\documentclass[12pt]{article}
\usepackage{graphicx}
\usepackage{lscape}
\usepackage{epsfig}
\usepackage{citesort}
\usepackage{amssymb}
\usepackage{amsmath}
\usepackage{multirow}
\usepackage[table]{xcolor}
\usepackage{colortbl}
\definecolor{lightgray}{gray}{0.9}

\newcommand{\be}{\begin{equation}}
\newcommand{\ee}{\end{equation}}
\newcommand{\bea}{\begin{eqnarray}}
\newcommand{\eea}{\end{eqnarray}}
\newcommand{\nn}{\nonumber}

\def\R1{\varepsilon_1}
\def\E8{\varepsilon_8}

\def\ga{\gamma}

\def\lb{\Lambda_b}

\def\s1{\hat s}
\def\ds{\displaystyle}

\newcommand{\bd}{\begin{displaymath}}
\newcommand{\ed}{\end{displaymath}}

\newcommand{\f}{\frac}

\def\R1{\varepsilon_1}
\def\E8{\varepsilon_8}

\def\ga{\gamma}

\def\ds{\displaystyle}
\def\beq{\begin{equation}}
\def\eeq{\end{equation}}
\def\bea{\begin{eqnarray}}
\def\eea{\end{eqnarray}}
\def\beeq{\begin{eqnarray}}
\def\eeeq{\end{eqnarray}}

\def\vel{\left|}
\def\ver{\right|}
\def\nnb{\nonumber}
\def\ga{\left(}
\def\dr{\right)}

\def\rar{\rightarrow}
\def\nnb{\nonumber}

\def\ba{\begin{array}}
\def\ea{\end{array}}

\def\xis0{{\Xi^{*0}}}

\def\g5{\gamma_5}

\def\es{\!\!\! &=& \!\!\!}

\def\ar{&+& \!\!\!}
\def\ek{&-& \!\!\!}

\newcommand{\al}{\alpha_s}

\begin{document}
\title{
         {\Large
                 {\bf Analysis of the semileptonic $\Lambda_b
\rightarrow \Lambda \ell^+ \ell^-$ transition
in topcolor-assisted technicolor (TC2) model
                 }
         }
      }
      
\author{\vspace{1cm}\\
{\small  K. Azizi$^1$ \thanks {e-mail: kazizi@dogus.edu.tr}\,\,, S. Kartal$^2$ \thanks
{e-mail: sehban@istanbul.edu.tr}\,\,, A. T. Olgun$^2$ \thanks
{e-mail: a.t.olgun@gmail.com}\,\,, Z. Tavuko\u glu$^2$ \thanks
{e-mail: z.tavukoglu@gmail.com}}  \\
{\small $^1$ Department of Physics, Do\u gu\c s University,
Ac{\i}badem-Kad{\i}k\"oy, 34722 \.{I}stanbul, Turkey}\\
{\small $^2$ Department of Physics, \.{I}stanbul University,
Vezneciler, 34134 \.{I}stanbul, Turkey}\\
}

\date{}
        \begin{titlepage}
\maketitle
\thispagestyle{empty}
\begin{abstract} 
We comparatively analyze the flavor changing neutral current process of the 
$\Lambda_b \rightarrow \Lambda \ell^+ \ell^-$ in the standard model as well as 
topcolor-assisted technicolor model using the form factors calculated via light cone QCD sum rules 
in full theory. In particular, we calculate the decay width, branching ratio and lepton forward-backward asymmetry
related to this decay channel. We compare the results of the topcolor-assisted 
technicolor model with those of the standard model and debate how the results of the 
topcolor-assisted technicolor model depart from the standard model predictions. We also compare our results
on the  branching ratio and differential branching ratio with recent experimental data provided by CDF and LHCb Collaborations.
\end{abstract}

~~~PACS number(s): 12.60.-i, 12.60.Nz, 13.30.-a, 13.30.Ce, 14.20.Mr
\end{titlepage}


\section{Introduction}  
The flavor changing neutral current (FCNC) processes in baryonic sector are promising tools in indirectly search  
for new physics (NP) effects besides the direct searches at large hadron colliders. Hence, both experimental and 
phenomenological works devoted to the analysis of these channels receive special attention nowadays. The 
semileptonic FCNC decay of $\Lambda_{b} \rightarrow \Lambda \ell^+ \ell^-$ ($\ell=e,\mu,\tau$) is 
one of the most important transitions in this respect since a heavy quark with a light di-quark combination
of the $\Lambda_{b}$ baryon makes this process significantly different than the B meson decays.
Experimentally, the CDF Collaboration at Fermilab has firstly reported their observation of the semileptonic 
$\Lambda_{b}^0 \rightarrow \Lambda \mu^+ \mu^-$ decay with a statistical 
significance of $5.8~\sigma$ and $24$ signal events at $\sqrt{s}=1.96$ TeV, collected by the CDF II detector 
in 2011 \cite{CDF}. They measured a branching ratio of 
$[1.73\pm 0.42(stat)\pm 0.55(syst)]\times 10^{-6}$ \cite{CDF} at muon channel. Recently, 
the LHCb Collaboration at CERN has also reported their observation of 
$\Lambda_{b}^0 \rightarrow \Lambda \mu^+ \mu^-$ with a signal yield of $78\pm12$, collected by 
the LHCb detector corresponding to an integrated luminosity of $1.0~ fb^{-1}$  at $\sqrt{s}=7$ TeV \cite{LHCb}.
They measured a branching ratio of $[0.96\pm 0.16(stat)\pm 0.13(syst)\pm 0.21(norm)]\times 10^{-6}$ at muon channel.
Until now, there is no direct evidence for the NP effects beyond the standard model (SM) 
at present particle physics experiments. However, the ATLAS and CMS Collaborations at CERN
reported their observations of a new particle like the SM Higgs boson with a mass of 
$\sim 125$ GeV at a statistical significance of $5~\sigma$ in 2012 \cite{ATLASColl,CMSColl}. Taken into 
consideration the above experimental progress and recent developments at LHC, with increasing the center of mass
energy, we hope we will be able to 
search for more FCNC decay processes as well as NP effects in near future. Therefore, theoretical 
calculations on NP effects to the FCNC processes using different scenarios will be required for analysis of 
the experimental results.  

In the present work, we analyze the semileptonic FCNC channel of 
the $\Lambda_{b} \rightarrow \Lambda \ell^+ \ell^-$ in the topcolor-assisted technicolor (TC2) model. We calculate
many parameters such as decay width, branching ratio and forward-backward asymmetry (FBA) and look for the difference 
between the results with those of the SM predictions. We also compare our results with the above mentioned
experimental data. 

The technicolor (TC) mechanism provides us with an alternative explanation for the origin of masses of
 the electroweak gauge bosons $W^{\pm}$ and $Z^{0}$ \cite{Kaul}. Although the TC and extended TC (ETC) models
 explain the flavor symmetry and electroweak symmetry
 breakings (EWSB), these models are unable to 
answer why mass of the top quark is too large \cite{Hill,Lane-Eichten}. In topcolor models, however, the top quark
involves in a new strong interaction, spontaneously broken at some high energy scales but not confining. 
According to these models, the strong dynamics provide the formation of top quark condensate $\overline{t}t$
which leads to a large dynamical mass for this quark, but with unnatural fine tuning \cite{Kominis,Lane-Eichten}.
Hence, a TC model containing “topcolor” scenario (TC2 model) has been developed \cite{Hill}. 
This model explains the electroweak and flavor symmetry breakings as well as the large mass of the top quark without
unnatural fine tuning. This model predicts the existence of top-pions ($\pi_{t}^{0,\pm}$),
the top-Higgs ($h_{t}^{0}$) and 
the non-universal gauge boson ($Z^{\prime}$), which we are going to discuss the dependences of the physical quantities
defining the $\Lambda_{b} \rightarrow \Lambda \ell^+ \ell^-$ transition on the masses of these objects in
this article. For more details of the TC2 model and some of its applications see 
for instance \cite{C7C8TC2,Xiao,YTC2ZTC2,HeffTC2} and references therein.

The outline of the article is as follows. In the next section, after briefly introducing the TC2 scenario
we present the effective Hamiltonian of the transition under consideration including Wilson coefficients as well as 
transition amplitude and matrix elements in terms of form factors. In section 3, we calculate
the differential decay width in TC2 model and numerically analyze the differential branching ratio, 
the branching ratio and the lepton forward-backward asymmetry and compare the obtained results with those of the SM
as well as existing experimental data.

\section{The semileptonic $\Lambda_b \rightarrow \Lambda \ell^+ \ell^-$ 
transition in SM and the topcolor-assisted technicolor models}       
In this section, first we give a brief overview on TC2 model. Then
we present the effective Hamiltonian in both SM and TC2 models and show how the 
Wilson coefficients entered to the low energy effective
Hamiltonian are changed in TC2 model compare to the SM. We also present the 
transition matrix elements in terms of form factors in full QCD and calculate the transition amplitude of the
$\Lambda_b \rightarrow \Lambda \ell^+ \ell^-$ in both models.
\subsection{The TC2 model}
As we previously mentioned, the TC2 model creates an attractive scheme because it combines the TC interaction
which is responsible for dynamical EWSB mechanism as an alternative to Higgs scenario as well as
topcolor interaction for the third generation at scale of near $\sim1$ TeV. 
This model provides an explanation of the electroweak and flavor symmetry breakings
 and also the large mass of top quark.  
  This model predicts a triplet of strongly coupled pseudo-Nambu-Goldstone bosons,
 neutral-charged “top-pions” ($\pi_{t}^{0,\pm}$) near the top mass scale, one isospin-singlet boson, 
the neutral “top-Higgs” ($h_t^0$) and the non-universal gauge boson ($Z^{\prime}$). Exchange of these
 new particles generates flavor-changing (FC) effects which lead to changes in the Wilson coefficients 
 compared to the SM \cite{Hill}.

The flavor-diagonal (FD) couplings of top-pions to the fermions are defined 
as \cite{Hill,Lane-Eichten,Lane,Hill-Simmons,Cvetic}
\begin{eqnarray}
&&\frac{m_t^*}{\sqrt{2}F_{\pi}} \frac{\sqrt{\nu_{w}^{2}-
F_{\pi}^{2}}}{\nu_{w}}\left[i\bar{t}\gamma^{5}t\pi_{t}^{0}+\sqrt{2}\bar{t}_{R}b_{L}
\pi_{t}^{+}+\sqrt{2}\bar{b}_{L}t_{R}\pi_{t}^{-}\right] \nonumber \\
&&+\frac{m_{b}^{*}}{\sqrt{2}F_{\pi}}\left[i\bar{b}\gamma^{5}b\pi_{t}^{0}+\sqrt{2}
\bar{t}_{L}b_{R}\pi_{t}^{+}+\sqrt{2}\bar{b}_{R}t_{L}\pi_{t}^{-}\right]
+\frac{m_{l}}{\nu}\bar{l}\gamma^{5}l\pi^{0}_{t},
\end{eqnarray}
where $m^*_{t}=m_{t}(1-\varepsilon)$ and $m_b^*=m_{b}-0.1~\varepsilon~m_{t}$ denote the masses of 
the top and bottom quarks generated by topcolor interactions, respectively. Here, $F_{\pi}$ is 
the physical top-pion decay constant which is estimated from the Pagels-Stokar formula, 
$\nu_{w}=\nu/\sqrt{2}=174\rm~GeV$, wherein the $\nu$ is defined as the vacuum expectation of the Higgs field 
and the factor $\frac{\sqrt{\nu_{w}^{2}-F_{\pi}^{2}}}{\nu_{w}}$ 
represents mixing effect between the Goldstone bosons (techni-pions) and top-pions.

The FC couplings of top-pions to the quarks can be written as \cite{BBHK,He-Yuan,GBurdman} 
\begin{eqnarray}
&&\frac{m_{t}^*}{\sqrt{2}F_{\pi}}\frac{\sqrt{\nu_{w}^{2}-
F_{\pi}^{2}}}{\nu_{w}}\left[iK_{UR}^{tc}K_{UL}^{tt^{*}}\bar{t}_{L}c_{R}\pi_{t}^{0}
+\sqrt{2}K_{UR}^{tc^{*}}K_{DL}^{bb}\bar{c}_{R}b_{L}\pi_{t}^{+}
+\sqrt{2}K_{UR}^{tc}K_{DL}^{bb^{*}}\bar{b}_{L}c_{R}\pi_{t}^{-}\right.\nonumber \\
&&\hspace{15mm}\left.+\sqrt{2}K_{UR}^{tc^{*}}K_{DL}^{ss}\bar{t}_{R}s_{L}\pi_{t}^{+}
+\sqrt{2}K_{UR}^{tc}K_{DL}^{ss^{*}}\bar{s}_{L}t_{R}\pi_{t}^{-}\right],
\end{eqnarray}
where $K_{UL(R)}$ and $K_{DL(R)}$ are rotation matrices that diagonalize the up-quark and 
down-quark mass matrices $M_{U}$ and $M_{D}$ for the down-type left and right hand quarks, 
respectively. The values of the coupling parameters are given as
\begin{equation}
K_{UL}^{tt}\approx K_{DL}^{bb} \approx K_{DL}^{ss}\approx1,
\hspace{10mm} K_{UR}^{tc}\leq\sqrt{2\varepsilon-\varepsilon^{2}} .
\end{equation}

The FD couplings of the new gauge boson $Z^\prime$ to the fermions are also given by 
\cite{Hill,Lane-Eichten,Lane,BBHK,Hill-Simmons,Cvetic}
\begin{eqnarray}
{\cal L}^{FD}_{Z'}&=&-\sqrt{4\pi K_{1}}\left\{
Z'_{\mu}\left[\frac{1}{2}\bar{\tau}_{L}\gamma^{\mu}\tau_{L}
-\bar{\tau}_{R}\gamma^{\mu}\tau_{R}+\frac{1}{6}\bar{t}_{L}\gamma^{\mu}t_{L}
+\frac{1}{6}\bar{b}_{L}\gamma^{\mu}b_{L}+\frac{2}{3}\bar{t}_{R}\gamma^{\mu}t_{R}
\right.\right.\nonumber\\
&-&\left.\left.\frac{1}{3}\bar{b}_{R}\gamma^{\mu}b_{R}\right]-\tan^{2}
\theta'Z'_{\mu}\left[\frac{1}{6}\bar{s}_{L}\gamma^{\mu}s_{L}
-\frac{1}{3}\bar{s}_{R}\gamma^{\mu}s_{R}-\frac{1}{2}
\bar{\mu}_{L}\gamma^{\mu}\mu_{L}-\bar{\mu}_{R}\gamma^{\mu}\mu_{R}
\right.\right.\nonumber\\
&-&\left.\left.\frac{1}{2}\bar{e}_{L}\gamma^{\mu}e_{L}-\bar{e}_{R}\gamma^{\mu}e_{R}\right]\right\},
\end{eqnarray} 
where $K_{1}$ is the coupling constant and it is taken in the region (0.3~-~1), $\theta'$ is the mixing angle and
 $tan\theta'=\frac{g_{1}}{\sqrt{4\pi K_1}}$ with  $g_{1}$ being the ordinary hypercharge
 gauge coupling constant.

The FC couplings of the non-universal  $Z^\prime$ gauge boson to the fermions can be written as 
\cite{CTHill}
\begin{eqnarray}
  {\cal L}^{FC}_{Z'}=-\frac{g_1}{2}\cot{\theta'} Z'^{\mu}\left\{\frac{1}{3}D_{L}^{bb} D_{L}^{bs*}
  \bar{s}_L\gamma_{\mu} b_L -
\frac{2}{3}D_{R}^{bb}D_{R}^{bs*}\bar{s}_R \gamma_{\mu} b_R  +{\rm
h.c.} \right\} \label{bsz},
\end{eqnarray}
where $D_L$ and $D_R$ are matrices rotate the weak eigen-basis to the mass ones 
for down-type left and right hand quarks, respectively. 
\subsection{The effective Hamiltonian and Wilson Coefficients}
The  $\Lambda_b \rightarrow \Lambda \ell^+ \ell^-$ decay is governed by $b \rightarrow s \ell^+ \ell^-$
transition at quark level in SM, whose effective Hamiltonian is given by \cite{Buchalla,Bobeth,Ball,Hurth}
\begin{eqnarray} \label{Heff} {\cal H}^{eff}_{SM} &=& \frac {G_F \alpha_{em} V_{tb}
V_{ts}^\ast}{2\sqrt{2} \pi} \Bigg[ C^{eff}_{9}
\bar{s}\gamma_\mu (1-\gamma_5) b \, \bar{\ell} \gamma^\mu \ell +
C_{10}  \bar{s} \gamma_\mu (1-\gamma_5) b \, \bar{\ell}
\gamma^\mu
\gamma_5 \ell \nnb \\
&-&  2 m_b C^{eff}_{7} \frac{1}{q^2} \bar{s} i \sigma_{\mu\nu} q^{\nu}
(1+\gamma_5) b \, \bar{\ell} \gamma^\mu \ell \Bigg]~,
\end{eqnarray}
where $G_{F}$ is the Fermi coupling constant, $\alpha_{em}$ is the fine structure constant, 
$V_{tb}$ and $V_{ts}^\ast$ are elements of the Cabibbo-Kobayashi-Maskawa (CKM) matrix,
 $q^{2}$ is the transferred momentum squared and the $C^{eff}_{7}$, $C^{eff}_{9}$, $C_{10}$ are 
the Wilson coefficients. Our main task in the following are to present the expressions of 
the Wilson coefficients.
The Wilson coefficient $C_7^{eff}$ in leading log approximation in SM is given 
by \cite{C7eff,Misiak,Muenz,Buras}
\begin{eqnarray}
\label{wilson-C7eff} C_7^{eff}(\mu_b) \es
\eta^{\frac{16}{23}} C_7(\mu_W)+ \frac{8}{3} \left(
\eta^{\frac{14}{23}} -\eta^{\frac{16}{23}} \right) C_8(\mu_W)+C_2 (\mu_W) 
\sum_{i=1}^8 h_i \eta^{a_i}~, \nnb\\ 
\end{eqnarray}
where
\begin{eqnarray} \eta \es
\frac{\alpha_s(\mu_W)} {\alpha_s(\mu_b)}~,
\end{eqnarray}
and
\begin{eqnarray}
\alpha_s(x)=\frac{\alpha_s(m_Z)}{1-\beta_0\frac{\alpha_s(m_Z)}{2\pi}\ln(\frac{m_Z}{x})}~,
\end{eqnarray}
with $\alpha_s(m_Z)=0.118$ and $\beta_0=\frac{23}{3}$. The remaining functions in
 Eq.\eqref{wilson-C7eff} are written as
\begin{eqnarray} 
C_7(\mu_W)=-\frac{1}{2}
D^{\prime~SM}_{0}(x_t)~,~~ C_8(\mu_W)=-\frac{1}{2}
E^{\prime~SM}_{0}(x_t)~,~~C_2(\mu_W)=1~, 
\end{eqnarray} 
where the functions $D^{\prime~SM}_{0}(x_t)$ and $E^{\prime~SM}_{0}(x_t)$ with
 $x_t=\frac{m_{t}^{2}}{m_{W}^{2}}$ are given by 
\begin{eqnarray} \label{Dprime0SM} 
D^{\prime~SM}_{0}(x_t) \es -
\frac{(8 x_t^3+5 x_t^2-7 x_t)}{12 (1-x_t)^3}
+ \frac{x_t^2(2-3 x_t)}{2(1-x_t)^4}\ln x_t~,
\end{eqnarray} 
and
\begin{eqnarray} \label{Eprime0SM} E^{\prime~SM}_{0}(x_t) \es - \frac{x_t(x_t^2-5 x_t-2)}
{4(1-x_t)^3} + \frac{3 x_t^2}{2 (1-x_t)^4}\ln x_t~. 
\end{eqnarray} 
The coefficients $h_i$ and $a_i$ in Eq.\eqref{wilson-C7eff}, with i running from 1 to 8, are also
given by \cite{Misiak,Muenz}
\be\frac{}{}
   \label{hi}
\begin{array}{rrrrrrrrrl}
h_i = (\!\! & 2.2996, & - 1.0880, & - \f{3}{7}, & - \f{1}{14}, &
-0.6494, & -0.0380, & -0.0186, & -0.0057 & \!\!)  \vspace{0.1cm}, \\
\end{array}
\ee 
and
\be\frac{}{}
   \label{ai}
\begin{array}{rrrrrrrrrl}
a_i = (\!\! & \f{14}{23}, & \f{16}{23}, & \f{6}{23}, & -
\f{12}{23}, &
0.4086, & -0.4230, & -0.8994, & 0.1456 & \!\!).
\end{array}
\ee 
The Wilson coefficient $C_9^{eff}$ in SM is expressed as \cite{Misiak,Muenz}
\begin{eqnarray} \label{wilson-C9eff}
C_9^{eff}(\hat{s}^{\prime}) & = & C_9^{NDR}\eta(\hat s^{\prime}) + h(z, \hat s^{\prime})\left( 3
C_1 + C_2 + 3 C_3 + C_4 + 3C_5 + C_6 \right) \nonumber \\
& & - \f{1}{2} h(1, \hat s^{\prime}) \left( 4 C_3 + 4 C_4 + 3C_5 + C_6 \right) \nonumber \\
& & - \f{1}{2} h(0, \hat s^{\prime}) \left( C_3 + 3 C_4 \right)
+ \f{2}{9} \left( 3 C_3 + C_4 + 3 C_5 +C_6 \right), 
\end{eqnarray}
where $\hat s^{\prime}=\frac{q^{2}}{m_{b}^{2}}$ with $q^2$ lies in the interval 
$4m_{l}^{2}\leq q^{2}\leq (m_{\Lambda_b}-m_{\Lambda})^{2}$.
The $C_9^{NDR}$ in the naive dimensional regularization (NDR) scheme is given as
\begin{eqnarray} \label{C9NDR}C_9^{NDR} & = & P_0^{NDR} +
\f{Y^{SM}}{\sin^2\theta_W} -4 Z^{SM} + P_E E^{SM}, 
\end{eqnarray} 
where $P_0^{NDR}=2.60 \pm 0.25$, $\sin^2\theta_W=0.23$, $Y^{SM}=0.98$ and $Z^{SM}=0.679$
 \cite{Misiak,Muenz,Buras}. The last term in Eq.\eqref{C9NDR} is ignored due to smallness of the
 value of $P_E$. The $\eta(\hat s^{\prime})$ in Eq.\eqref{wilson-C9eff} is also defined as
\begin{eqnarray} \eta(\hat s^{\prime}) & = & 1 + \f{\al(\mu_b)}{\pi}\,
\omega(\hat s^{\prime}), 
\end{eqnarray} 
with
\begin{eqnarray} \label{omega-shat}
\omega(\hat s^{\prime}) & = & - \f{2}{9} \pi^2 - \f{4}{3}\mbox{Li}_2(\hat s^{\prime}) - \f{2}{3}
\ln \hat s^{\prime} \ln(1-\hat s^{\prime}) - \f{5+4\hat s^{\prime}}{3(1+2\hat s^{\prime})}
\ln(1-\hat s^{\prime}) - \nonumber \\
& &  \f{2 \hat s^{\prime} (1+\hat s^{\prime}) (1-2\hat s^{\prime})}{3(1-\hat s^{\prime})^2
(1+2\hat s^{\prime})} \ln \hat s^{\prime} + \f{5+9\hat s^{\prime}-6\hat s^{\prime2}}
{6 (1-\hat s^{\prime}) (1+2\hat s^{\prime})}. 
\end{eqnarray}
The function $h(y,\hat s^{\prime})$ is given as
\begin{eqnarray} \label{h-phasespace} h(y,
\hat s^{\prime}) & = & -\f{8}{9}\ln\f{m_b}{\mu_b} - \f{8}{9}\ln y +
\f{8}{27} + \f{4}{9} x \\
& & - \f{2}{9} (2+x) |1-x|^{1/2} \left\{
\begin{array}{ll}
\left( \ln\left| \f{\sqrt{1-x} + 1}{\sqrt{1-x} - 1}\right| - i\pi
\right), &
\mbox{for } x \equiv \f{4z^2}{\hat s^{\prime}} < 1 \nonumber \\
2 \arctan \f{1}{\sqrt{x-1}}, & \mbox{for } x \equiv \f {4z^2}{\hat
s^{\prime}} > 1,
\end{array}
\right. \\
\end{eqnarray}
where 
$y=1$ or $y=z=\frac{m_c}{m_b}$ 
and 
\begin{eqnarray} h(0, \hat s^{\prime})
& = & \f{8}{27} -\f{8}{9} \ln\f{m_b}{\mu_b} - \f{4}{9} \ln \hat s^{\prime}
+ \f{4}{9} i\pi.
\end{eqnarray}
At $\mu_b=5~GeV$ scale, the coefficients $C_j$ (j=1,...6) are given by \cite{Buras}
\begin{eqnarray} \label{CJ} C_j=\sum_{i=1}^8 k_{ji}
\eta^{a_i} \qquad (j=1,...6) \vspace{0.2cm}, 
\end{eqnarray}
where the constants $k_{ji}$ are given as \be\frac{}{}
   \label{KJI}
\begin{array}{rrrrrrrrrl}
k_{1i} = (\!\! & 0, & 0, & \f{1}{2}, & - \f{1}{2}, &
0, & 0, & 0, & 0 & \!\!),  \vspace{0.1cm} \\
k_{2i} = (\!\! & 0, & 0, & \f{1}{2}, &  \f{1}{2}, &
0, & 0, & 0, & 0 & \!\!),  \vspace{0.1cm} \\
k_{3i} = (\!\! & 0, & 0, & - \f{1}{14}, &  \f{1}{6}, &
0.0510, & - 0.1403, & - 0.0113, & 0.0054 & \!\!),  \vspace{0.1cm} \\
k_{4i} = (\!\! & 0, & 0, & - \f{1}{14}, &  - \f{1}{6}, &
0.0984, & 0.1214, & 0.0156, & 0.0026 & \!\!),  \vspace{0.1cm} \\
k_{5i} = (\!\! & 0, & 0, & 0, &  0, &
- 0.0397, & 0.0117, & - 0.0025, & 0.0304 & \!\!) , \vspace{0.1cm} \\
k_{6i} = (\!\! & 0, & 0, & 0, &  0, &
0.0335, & 0.0239, & - 0.0462, & -0.0112 & \!\!).  \vspace{0.1cm} \\
\end{array}
\ee
The Wilson coefficient $C_{10}$ in SM is scale-independent and has the following 
explicit expression:
\begin{eqnarray} \label{wilson-C10} 
C_{10}= - \frac{Y^{SM}}{\sin^2 \theta_W}~. 
\end{eqnarray}

The effective Hamiltonian for $b \rightarrow s \ell^+ \ell^-$ transition in 
TC2 model is given by \cite{HeffTC2}
\begin{eqnarray} \label{HeffTC2} 
{\cal H}^{eff}_{TC2} &=& \frac{G_F \alpha_{em} V_{tb}
V_{ts}^\ast}{2\sqrt{2} \pi} \Bigg[ \widetilde{C}^{eff}_{9}
\bar{s}\gamma_\mu (1-\gamma_5) b \, \bar{\ell} \gamma^\mu \ell +
\widetilde{C}_{10}  \bar{s} \gamma_\mu (1-\gamma_5) b \, \bar{\ell}
\gamma^\mu
\gamma_5 \ell \nnb \\
&-&  2 m_b \widetilde{C}^{eff}_{7} \frac{1}{q^2} \bar{s} i \sigma_{\mu\nu} q^{\nu}
(1+\gamma_5) b \, \bar{\ell} \gamma^\mu \ell + C_{Q_1}\bar{s}(1+\gamma_5)b\bar{\ell}\ell \nnb \\
&+& C_{Q_2}\bar{s}(1+\gamma_5)b\bar{\ell}\gamma_5\ell\Bigg]~,
\end{eqnarray}  
where $\widetilde{C}^{eff}_{7}$, $\widetilde{C}^{eff}_{9}$, $\widetilde{C}_{10}$, $C_{Q_1}$ and $C_{Q_2}$
are new Wilson coefficients. The $\widetilde{C}^{eff}_{7}$, $\widetilde{C}^{eff}_{9}$ and $\widetilde{C}_{10}$ 
contain contributions from both the SM and TC2 models. The charged top-pions 
$\pi_{t}^{\pm}$ only give contributions to the Wilson coefficient $\widetilde{C}^{eff}_{7}$ while 
the non-universal gauge boson $Z^\prime$ contributes to the Wilson coefficients  
$\widetilde{C}^{eff}_{9}$ and $\widetilde{C}_{10}$. In the following, we present the explicit expressions 
of the Wilson coefficients $C_7(\mu_W)$ and $C_8(\mu_W)$ 
entered to Eq.\eqref{wilson-C7eff} in TC2 model. The new photonic-and gluonic-penguin diagrams in TC2 model
can be gotten by replacement of the internal $W^{\pm}$ lines in SM penguin diagrams with unit-charged scalar 
($\pi_{1}^{\pm}$, $\pi_{8}^{\pm}$ and $\pi_{t}^{\pm}$) lines 
(for more information, see Ref. \cite{C7C8TC2}). As a result, the Wilson coefficients
 $\widetilde{C}_{7}(\mu_W)$ and $\widetilde{C}_{8}(\mu_W)$ in TC2 take the following 
forms \cite{HeffTC2,D0pTC2}:
\begin{eqnarray} \widetilde{C}_{7}(\mu_W)=-\frac{1}{2}
D^{\prime~TC2tot}_{0}(x_t,z_t)~,~~ \widetilde{C}_{8}(\mu_W)=-\frac{1}{2}
E^{\prime~TC2tot}_{0}(x_t,z_t)~, 
\end{eqnarray} 
where
\begin{eqnarray} D^{\prime~TC2tot}_{0}(x_t,z_t)&=&D_{0}^{\prime~SM}(x_t)+D_{0}^{\prime~TC2}(z_t),\nnb \\
E^{\prime~TC2tot}_{0}(x_t,z_t)&=&E_{0}^{\prime~SM}(x_t)+E_{0}^{\prime~TC2}(z_t),
\end{eqnarray} 
and 
\begin{eqnarray}
D_{0}^{\prime~TC2}(z_t)&=&\frac{1}{8\sqrt{2}G_FF_{\pi}^2}\left[-\frac{22-53z_t+25z_t^2}
{18(1-z_t)^3}-\frac{3z_t-8z_t^2+4z_t^3}{3(1-z_t)^4}{\rm log}[z_t]\right],\nnb \\
E_{0}^{\prime~TC2}(z_t)&=&\frac{1}{8\sqrt{2}G_FF_{\pi}^2}\left[-\frac{5-19z_t+20z_t^2}
{6(1-z_t)^3}+\frac{z_t^2-2z_t^3}{(1-z_t)^4}{\rm log}[z_t]\right],
\end{eqnarray} 
with $z_t={m^*_t}^2/m_{\pi_t^{\pm}}^2$.\\
There are new contributions coming from the non-universal gauge boson $Z^{\prime}$ in TC2 model 
to the $Y^{SM}$ and $Z^{SM}$ entered to the ${C}_{9}^{NDR}$ in Eq.\eqref{wilson-C9eff} \cite{HeffTC2}.
 The $\widetilde{C}_{9}^{NDR}$ in TC2 model is given by 
\begin{eqnarray}
\widetilde{C}_{9}^{NDR} = P_{0}^{NDR} + \frac{Y^{TC2tot}(y_t)}{\sin^2\theta_W} -4 Z^{TC2tot}(y_t),
\end{eqnarray}  
where 
\begin{eqnarray}
Y^{TC2tot}(y_t)&=&Y^{SM}+Y^{TC2}(y_t), \nnb \\
Z^{TC2tot}(y_t)&=&Z^{SM}+Z^{TC2}(y_t).       
\end{eqnarray}    
 The functions $Y^{TC2}(y_t)$ and $Z^{TC2}(y_t)$ are given by the following expressions in the case of $e$ or $\mu$
 in the final state \cite{YTC2ZTC2,HeffTC2}:
\begin{eqnarray} \label{YTC2}
Y^{TC2}(y_t)=Z^{TC2}(y_t)&=&\frac{-tan^2\theta'M_Z^2}{M_{Z'}^2} \Big[K_{ab}(y_t)+K_{c}(y_t)+K_{d}(y_t) \Big],
\end{eqnarray}
with $y_t=m_{t}^{*2}/m_{W}^2$. In the case of $\tau$ as final leptonic state, the factor $-tan^2\theta'$ in the above
equation is replaced by 1. The functions in Eq.\eqref{YTC2} are also defined as \cite{Kabcd}
\begin{eqnarray}
K_{ab}(y_t)&=&\frac{8}{3}(tan^2\theta'-1)\frac{F_1(y_t)}{(v_d+a_d)},\nnb \\
K_{c}(y_t)&=&\frac{16F_2(y_t)}{3(v_u-a_u)}-\frac{8F_3(y_t)}{3(v_u+a_u)},\nnb \\
K_{d}(y_t)&=&\frac{16F_4(y_t)}{3(v_u-a_u)}+\frac{8F_5(y_t)}{3(v_u+a_u)},
\end{eqnarray}
where  
\begin{eqnarray}   
F_1(y_t)&=&-\Big[0.5(Q-1)\sin^2\theta_W+0.25 \Big]\Bigg\{y_t^2 ln(y_t)/(y_t-1)^2-y_t/(y_t-1) \nnb \\
&-& y_t\Big[0.5\Big(-0.5772+ln(4\pi)-ln(M_W^2)\Big)+0.75-0.5\Big(x^2ln(y_t)/(y_t-1)^2-1/(y_t-1)\Big)\Big]\Bigg\}\nnb \\
&& \Big[(1+y_t)/(y_t-2)\Big],\nnb \\
F_2(y_t)&=&\Big(0.5 Q \sin^2\theta_W-0.25\Big)\Big[y_t^2ln(y_t)/(y_t-1)^2-2y_tln(y_t)/
(y_t-1)^2+y_t/(y_t-1)\Big],\nnb \\
F_3(y_t)&=&-Q \sin^2\theta_W \Big[y_t/(y_t-1)-y_tln(y_t)/(y_t-1)^2\Big],\nnb \\
F_4(y_t)&=&0.25\Big(4 \sin^2\theta_W/3-1\Big) \Big[y_t^2ln(y_t)/(y_t-1)^2-y_t-y_t/(y_t-1)\Big],\nnb \\
F_5(y_t)&=&-0.25Q \sin^2\theta_W y_t\Big[-0.5772+ln(4\pi)-ln(M_W^2)+1-y_tln(y_t)/(y_t-1)\Big] \nnb \\
&-& \sin^2\theta_W/6 \Big[y_t^2ln(y_t)/(y_t-1)^2-y_t-y_t/(y_t-1)\Big],    
\end{eqnarray}  
and $a_{u,d}=I_3$, $v_{u,d}=I_3-2Q_{u,d}\sin^2\theta_W$ with $u$ and $d$ representing the up and down type
 quarks, respectively.\\ 
The Wilson coefficients $C_{Q_1}$ and $C_{Q_2}$ appeared in the effective Hamiltonian in TC2 model belong to
the neutral top-pion $\pi_{t}^{0}$ and top-Higgs $h_{t}^{0}$ contributing to the rare decays.
 The coefficient $C_{Q_1}$ in TC2 model is given by \cite{YTC2ZTC2}
\begin{eqnarray} \label{CQ1} 
C_{Q_1}  = \frac{\sqrt{\nu_w^2-F_{\pi}^2}}{\nu_w}\left(
\frac{m_b^*m_l\nu} {2\sqrt{2}sin^2\theta_wF_{\pi}m_{h_t^0}^2
}C_0(x_t)+\frac{V_{ts}m_lm_tm_b^{*2}M_W^2 }{4\sqrt{2} g_2^4
F_{\pi}^2m_{h_t^0}^2 }C(x_s)\right),
\end{eqnarray}
where the Inami-Lim function $C_0(x_t)$ is defined by \cite{Buras}
\begin{eqnarray}
C_0(x_t)&=&\frac{x_t}{8}\left[\frac{x_t-6}{x_t-1}+\frac{3x_t+2}{(x_t-1)^2}\ln x_t\right].
\end{eqnarray}
The $C(x_s)$ function in Eq.\eqref{CQ1} is also given as \cite{HeffTC2}
\begin{eqnarray}
C(x_s)&=&\frac{F_6(x)}{-[0.5(Q-1)\sin^2\theta_W+0.25]},
\end{eqnarray}
with
\begin{eqnarray}
F_6(x_s)&=&-\Big[0.5(Q-1)\sin^2\theta_W+0.25\Big] \Bigg\{x_s^2 ln(x_s)/(x_s-1)^2-x_s/(x_s-1)\nnb \\
&-& x_s\Big[0.5\Big(-0.5772+ln(4\pi)-ln(M_W^2)\Big)+0.75-0.5\Big(x^2ln(x_s)/(x_s-1)^2 \nnb \\
&-& 1/(x_s-1)\Big)\Big] \Bigg\},
\end{eqnarray}
where $x_s={m^*_t}^2/m_{\pi_t^0}^2$ and the $g_2$ is the $SU(2)$ coupling constant. 
Since, the neutral top-Higgs coupling with fermions differ from that of neutral top-pion by a factor
of $\gamma_5$, the form of $C_{Q_2}$ is the same as $C_{Q_1}$ except for the masses of 
the scalar particles \cite{YTC2ZTC2}.
\subsection{Transition amplitude and  matrix elements}
The transition amplitude for $\Lambda_b \rightarrow \Lambda \ell^+ \ell^-$ decay is 
obtained by sandwiching the effective Hamiltonian between the initial and final baryonic states
\begin{eqnarray}\label{amplitude}
{\cal M}^{\Lambda_b \rightarrow \Lambda \ell^+ \ell^-} = \langle \Lambda(p_{\Lambda})
 \mid{\cal H}^{eff}\mid \Lambda_b(p_{\Lambda_b})\rangle,
\end{eqnarray}
where $p_{\Lambda_b}$ and $p_{\Lambda}$ are momenta of the $\Lambda_b$ and $\Lambda$ baryons, 
respectively. In order to calculate the transition amplitude, we need to define the following 
matrix elements in terms of twelve form factors in full QCD:
\begin{eqnarray}\label{SMtransmatrix1} \langle
\Lambda(p_{\Lambda}) \mid  \bar s \gamma_\mu (1-\gamma_5) b \mid \Lambda_b(p_{\Lambda_b})\rangle\es
\bar {u}_\Lambda(p_{\Lambda}) \Bigg[\gamma_{\mu}f_{1}(q^{2})+{i}
\sigma_{\mu\nu}q^{\nu}f_{2}(q^{2}) + q^{\mu}f_{3}(q^{2}) \nnb \\
\ek \gamma_{\mu}\gamma_5
g_{1}(q^{2})-{i}\sigma_{\mu\nu}\gamma_5q^{\nu}g_{2}(q^{2})
- q^{\mu}\gamma_5 g_{3}(q^{2})
\vphantom{\int_0^{x_2}}\Bigg] u_{\Lambda_{b}}(p_{\Lambda_b})~,\nnb \\
\end{eqnarray}
\begin{eqnarray}\label{SMtransmatrix2}
\langle \Lambda(p_{\Lambda})\mid \bar s i \sigma_{\mu\nu}q^{\nu} (1+ \gamma_5)
b \mid \Lambda_b(p_{\Lambda_b})\rangle \es\bar{u}_\Lambda(p_{\Lambda})
\Bigg[\gamma_{\mu}f_{1}^{T}(q^{2})+{i}\sigma_{\mu\nu}q^{\nu}f_{2}^{T}(q^{2})+
q^{\mu}f_{3}^{T}(q^{2}) \nnb \\
\ar \gamma_{\mu}\gamma_5
g_{1}^{T}(q^{2})+{i}\sigma_{\mu\nu}\gamma_5q^{\nu}g_{2}^{T}(q^{2})
+ q^{\mu}\gamma_5 g_{3}^{T}(q^{2})
\vphantom{\int_0^{x_2}}\Bigg] u_{\Lambda_{b}}(p_{\Lambda_b})~,\nnb \\
\end{eqnarray}
and
\begin{eqnarray}\label{TC2transmatrix1}\langle
\Lambda(p_{\Lambda}) \mid  \bar s (1+\gamma_5) b \mid \Lambda_b(p_{\Lambda_b})\rangle\es
\frac{1}{m_b} \bar {u}_\Lambda(p_{\Lambda}) \Bigg[{\not\!q}f_{1}(q^{2})+{i}
q^{\mu} \sigma_{\mu\nu}q^{\nu}f_{2}(q^{2}) + q^{2}f_{3}(q^{2}) \nnb \\
\ek {\not\!q}\gamma_5
g_{1}(q^{2})-{i}q^{\mu}\sigma_{\mu\nu}\gamma_5q^{\nu}g_{2}(q^{2})
- q^{2}\gamma_5 g_{3}(q^{2})
\vphantom{\int_0^{x_2}}\Bigg] u_{\Lambda_{b}}(p_{\Lambda_b})~,\nnb \\
\end{eqnarray}
where the $u_{\Lambda_b}$ and $u_{\Lambda}$ are spinors of the initial and final baryons. 
By the way, the $f^{(T)}_i$ and $g^{(T)}_i$ with $i=1,2$ and $3$ are transition form factors.

Using the above transition matrix elements, we find the transition 
amplitude for $\Lambda_b \rightarrow \Lambda \ell^+ \ell^-$ in TC2 model as
\begin{eqnarray}\label{amplitudeTC2}
&&{\cal M}_{TC2}^{ \Lambda_b \rightarrow \Lambda \ell^+ \ell^-} = 
\frac{G_F \alpha_{em} V_{tb}V_{ts}^\ast}{2\sqrt{2} \pi} \Bigg\{\nnb \\
&&\Big[{\bar u}_\Lambda ({p}_{\Lambda}) ( \gamma_{\mu}[{\cal A}_1 R + {\cal B}_1 L]+
 {i}\sigma_{\mu\nu} q^{\nu}[{\cal A}_2 R + {\cal B}_2 L] + q^{\mu} [{\cal A}_3 R +
 {\cal B}_3 L]) u_{\Lambda_{b}}(p_{\Lambda_b}) \Big] \, (\bar{\ell} \gamma^\mu \ell)\nnb \\
&+& \Big[{\bar u}_\Lambda ({p}_{\Lambda})( \gamma_{\mu}[{\cal D}_1 R + {\cal E}_1 L]+
 {i}\sigma_{\mu\nu} q^{\nu}[{\cal D}_2 R +{\cal E}_2 L] + q^{\mu} [{\cal D}_3 R + {\cal E}_3 L]) 
u_{\Lambda_{b}}(p_{\Lambda_b}) \Big] \,(\bar{\ell} \gamma^\mu \gamma_5 \ell) \nnb \\
&+& \Big[{\bar u}_\Lambda ({p}_{\Lambda})( {\not\!q} [{\cal G}_1 R + {\cal H}_1 L]+ {i} q^{\mu}
 \sigma_{\mu\nu} q^{\nu}[{\cal G}_2 R + {\cal H}_2 L] + q^2 [{\cal G}_3 R + {\cal H}_3 L])
 u_{\Lambda_{b}}(p_{\Lambda_b}) \Big] \, (\bar{\ell} \ell)\nnb \\
&+& \Big[{\bar u}_\Lambda ({p}_{\Lambda})( {\not\!q}[{\cal K}_1 R + {\cal S}_1 L]+ {i} q^{\mu} 
\sigma_{\mu\nu}q^{\nu}[{\cal K}_2 R + {\cal S}_2 L] + q^2 [{\cal K}_3 R + {\cal S}_3 L]) 
u_{\Lambda_{b}}(p_{\Lambda_b}) \Big] \, (\bar{\ell} \gamma_5 \ell) \Bigg\} ~,\nnb \\
\end{eqnarray}
where $R=(1+\gamma_5)/2$ and $L=(1-\gamma_5)/2$. The calligraphic coefficients are 
defined as
\begin{eqnarray} \label{coef-decay-rate} {\cal A}_{1} \es (f_1 - g_1)~\widetilde{C}_{9}^{eff}~-~2 m_b 
\frac{1}{q^2}~(f_1^T + g_1^T)~\widetilde{C}_{7}^{eff}~,~~~~~~~~~~{\cal G}_{1} = 
\frac{1}{m_b}~(f_1 - g_1)~C_{Q_1}~,~\nnb 
\end{eqnarray}
\begin{eqnarray}
{\cal A}_{2} = {\cal A}_1 ( 1 \rar 2 )~,~~~~~~~~~~~~~~~~~~~~~~~~~~~~~~~~~~~~~~~~~~~~~~~~~~ 
{\cal G}_{2} = {\cal G}_1 \ga 1 \rar 2 \dr~,~~~~\nnb 
\end{eqnarray}
\begin{eqnarray}
{\cal A}_{3} = {\cal A}_1 \ga 1 \rar 3 \dr~,~~~~~~~~~~~~~~~~~~~~~~~~~~~~~~~~~~~~~~~~~~~~~~~~~~ 
{\cal G}_{3} = {\cal G}_1 \ga 1 \rar 3 \dr~,~~~~\nnb 
\end{eqnarray}
\begin{eqnarray}
{\cal B}_{1} \es {\cal A}_1 ( g_{1} \rar -g_{1}~;~g_1^T \rar -g_1^T)~,~~~~~~~~~~~~~~~~~~~~~~~~~~~~~ 
{\cal H}_{1} = {\cal G}_1 ( g_{1} \rar -g_{1})~,~~~~~\nnb 
\end{eqnarray}
\begin{eqnarray}
{\cal B}_{2} = {\cal B}_1 \ga 1 \rar 2 \dr~,~~~~~~~~~~~~~~~~~~~~~~~~~~~~~~~~~~~~~~~~~~~~~~~~~~ 
{\cal H}_{2} = {\cal H}_1 \ga 1 \rar 2 \dr~,~~~\nnb 
\end{eqnarray}
\begin{eqnarray}
{\cal B}_{3} = {\cal B}_1 \ga 1 \rar 3 \dr~,~~~~~~~~~~~~~~~~~~~~~~~~~~~~~~~~~~~~~~~~~~~~~~~~~~
 {\cal H}_{3} =  {\cal H}_1 \ga 1 \rar 3 \dr~,~~~\nnb
\end{eqnarray}
\begin{eqnarray}
{\cal D}_{1} \es (f_1 - g_1)~\widetilde{C}_{10}~,~~~~~~~~~~~~~~~~~~~~~~~~~~~~~~~~~~~~~~~~~~~~~~~ 
{\cal K}_{1} = \frac{1}{m_b}~(f_1 - g_1)~C_{Q_2}~,~\nnb
\end{eqnarray}
\begin{eqnarray}
{\cal D}_{2} =  {\cal D}_1 \ga 1 \rar 2 \dr~,~~~~~~~~~~~~~~~~~~~~~~~~~~~~~~~~~~~~~~~~~~~~~~~~~~ 
{\cal K}_{2} =  {\cal K}_1 \ga 1 \rar 2 \dr~,~~~~\nnb
\end{eqnarray}
\begin{eqnarray}
{\cal D}_{3} = {\cal D}_1 \ga 1 \rar 3 \dr~,~~~~~~~~~~~~~~~~~~~~~~~~~~~~~~~~~~~~~~~~~~~~~~~~~~ 
{\cal K}_{3} =  {\cal K}_1 \ga 1 \rar 3 \dr~,~~~~~\nnb
\end{eqnarray}
\begin{eqnarray}
{\cal E}_{1} = {\cal D}_1 ( g_{1} \rar -g_{1})~,~~~~~~~~~~~~~~~~~~~~~~~~~~~~~~~~~~~~~~~~~~~~~~~ 
{\cal S}_{1} = {\cal K}_1 (g_{1} \rar -g_{1})~,~\nnb
\end{eqnarray}
\begin{eqnarray}
{\cal E}_{2} = {\cal E}_1 \ga 1 \rar 2 \dr~,~~~~~~~~~~~~~~~~~~~~~~~~~~~~~~~~~~~~~~~~~~~~~~~~~~~~
 {\cal S}_{2} = {\cal S}_1 \ga 1 \rar 2 \dr~,~~~~~\nnb
\end{eqnarray}
\begin{eqnarray}
{\cal E}_{3} = {\cal E}_1 \ga 1 \rar 3 \dr~,~~~~~~~~~~~~~~~~~~~~~~~~~~~~~~~~~~~~~~~~~~~~~~~~~~~~ 
{\cal S}_{3} = {\cal S}_1 \ga 1 \rar 3 \dr~.~~~~~\nnb\\
\end{eqnarray}
\section{Physical Observables}
In this section, we calculate some physical observables such as differential decay width, 
 differential branching ratio,  branching ratio and the lepton forward-backward asymmetry
 for the decay channel under consideration. 
\subsection{The differential decay width}
In the present subsection, using the above-mentioned amplitude, we find the differential decay rate
 in terms of form factors in full theory in TC2 model as 
\begin{eqnarray}\label{DDR} \frac{d^2\Gamma}{d\hat
sdz}(z,\hat s) = \frac{G_F^2\alpha^2_{em} m_{\Lambda_b}}{16384
\pi^5}| V_{tb}V_{ts}^*|^2 v \sqrt{\lambda(1,r,\hat s)} \, \Bigg[{\cal
T}_{0}(\hat s)+{\cal T}_{1}(\hat s) z +{\cal T}_{2}(\hat s)
z^2\Bigg]~, \nnb\\ \label{dif-decay}
\end{eqnarray}
where $v=\sqrt{1-\frac{4 m_\ell^2}{q^2}}$ is the lepton velocity, 
$\lambda=\lambda(1,r,\hat s)=(1-r-\hat s)^2-4r\hat s$ is the usual triangle function 
with $\hat s= q^2/m^2_{\Lambda_b}$, $r= m^2_{\Lambda}/m^2_{\Lambda_b}$, $z=\cos\theta$ and
 $\theta$ is the angle between momenta of the lepton $l^+$ and the  $\Lambda_b$ in the center of 
mass of leptons.
The calligraphic, ${\cal T}_{0}(\hat s)$, ${\cal T}_{1}(\hat s)$ and ${\cal T}_{2}(\hat s)$ functions
 are obtained as
 \begin{eqnarray} {\cal T}_{0}(\hat s) \es 32 m_\ell^2
m_{\Lambda_b}^4 \hat s (1+r-\hat s) \Big( \vel {\cal D}_{3} \ver^2 +
\vel {\cal E}_{3} \ver^2 \Big) \nnb \\
\ar 64 m_\ell^2 m_{\Lambda_b}^3 (1-r-\hat s) \, \mbox{\rm Re} \Big[{\cal D}_{1}^\ast
{\cal E}_{3} + {\cal D}_{3}
{\cal E}_1^\ast \Big] \nnb \\
\ar 64 m_{\Lambda_b}^2 \sqrt{r} (6 m_\ell^2 - m_{\Lambda_b}^2 \hat s)
{\rm Re} \Big[{\cal D}_{1}^\ast {\cal E}_{1}\Big] \nnb\\ 
\ar 64 m_\ell^2 m_{\Lambda_b}^3 \sqrt{r} \Bigg\{ 2 m_{\Lambda_b} \hat s
{\rm Re} \Big[{\cal D}_{3}^\ast {\cal E}_{3}\Big] + (1 - r + \hat s)
{\rm Re} \Big[{\cal D}_{1}^\ast {\cal D}_{3} + {\cal E}_{1}^\ast {\cal E}_{3}\Big]\Bigg\} \nnb \\
\ar 32 m_{\Lambda_b}^2 (2 m_\ell^2 + m_{\Lambda_b}^2 \hat s) \Bigg\{ (1
- r + \hat s) m_{\Lambda_b} \sqrt{r} \,
\mbox{\rm Re} \Big[{\cal A}_{1}^\ast {\cal A}_{2} + {\cal B}_{1}^\ast {\cal B}_{2}\Big] \nnb \\
\ek m_{\Lambda_b} (1 - r - \hat s) \, \mbox{\rm Re} \Big[{\cal A}_{1}^\ast {\cal B}_{2} +
{\cal A}_{2}^\ast {\cal B}_{1}\Big] - 2 \sqrt{r} \Big( \mbox{\rm Re} \Big[{\cal A}_{1}^\ast
 {\cal B}_{1}\Big] +
m_{\Lambda_b}^2 \hat s \,
\mbox{\rm Re} \Big[{\cal A}_{2}^\ast {\cal B}_{2}\Big] \Big) \Bigg\} \nnb \\
\ar 8 m_{\Lambda_b}^2 \Bigg\{ 4 m_\ell^2 (1 + r - \hat s) +
m_{\Lambda_b}^2 \Big[(1-r)^2 - \hat s^2 \Big]
\Bigg\} \Big( \vel {\cal A}_{1} \ver^2 +  \vel {\cal B}_{1} \ver^2 \Big) \nnb \\
\ar 8 m_{\Lambda_b}^4 \Bigg\{ 4 m_\ell^2 \Big[ \lambda + (1 + r -
\hat s) \hat s \Big] + m_{\Lambda_b}^2 \hat s \Big[(1-r)^2 - \hat s^2 \Big]
\Bigg\} \Big( \vel {\cal A}_{2} \ver^2 +  \vel {\cal B}_{2} \ver^2 \Big) \nnb \\
\ek 8 m_{\Lambda_b}^2 \Bigg\{ 4 m_\ell^2 (1 + r - \hat s) -
m_{\Lambda_b}^2 \Big[(1-r)^2 - \hat s^2 \Big]
\Bigg\} \Big( \vel {\cal D}_{1} \ver^2 +  \vel {\cal E}_{1} \ver^2 \Big) \nnb\
\end{eqnarray}
\begin{eqnarray}
\ar 8 m_{\Lambda_b}^5 \hat s v^2 \Bigg\{ - 8 m_{\Lambda_b} \hat s \sqrt{r}\,
\mbox{\rm Re} \Big[{\cal D}_{2}^\ast {\cal E}_{2}\Big] +
4 (1 - r + \hat s) \sqrt{r} \, \mbox{\rm Re}\Big[{\cal D}_{1}^\ast {\cal D}_{2}+{\cal E}_{1}^\ast 
{\cal E}_{2}\Big]\nnb \\
\ek 4 (1 - r - \hat s) \, \mbox{\rm Re}\Big[{\cal D}_{1}^\ast {\cal E}_{2}+{\cal D}_{2}^\ast 
{\cal E}_{1}\Big] +
m_{\Lambda_b} \Big[(1-r)^2 -\hat s^2\Big] \Big( \vel {\cal D}_{2} \ver^2 + \vel
{\cal E}_{2} \ver^2 \Big) \Bigg\} \nnb \\
\ek 8 m_{\Lambda_b}^4 \Bigg\{ 4 m_\ell \Big[(1-r)^2 -\hat s(1+r) \Big]\, \mbox{\rm Re} 
\Big[{\cal D}_{1}^\ast {\cal K}_{1}+{\cal E}_{1}^\ast {\cal S}_{1}\Big] \nnb \\ 
\ar (4 m_\ell^2 - m_{\Lambda_b}^2 \hat s) \Big[(1-r)^2 -\hat s(1+r) \Big]\, 
\Big( \vel {\cal G}_{1} \ver^2 + \vel {\cal H}_{1} \ver^2 \Big) \nnb \\
\ar 4 m_{\Lambda_b}^2 \sqrt{r} \hat s^2 (4 m_\ell^2 - m_{\Lambda_b}^2 \hat s) \, 
\mbox{\rm Re}\Big[{\cal G}_{3}^\ast {\cal H}_{3}\Big] \Bigg\} \nnb \\
\ek 8 m_{\Lambda_b}^5 \hat s \Bigg\{2 \sqrt{r} (4 m_\ell^2 - m_{\Lambda_b}^2 \hat s) \,
(1 - r + \hat s) \, \mbox{\rm Re}\Big[{\cal G}_{1}^\ast {\cal G}_{3}+{\cal H}_{1}^\ast 
{\cal H}_{3}\Big] \nnb \\
\ar 4 m_\ell \sqrt{r}(1 - r + \hat s) \mbox{\rm Re}\Big[{\cal D}_{1}^\ast
 {\cal K}_{3}+{\cal E}_{1}^\ast {\cal S}_{3}+{\cal D}_{3}^\ast {\cal K}_{1}+{\cal E}_{3}^\ast 
{\cal S}_{1}\Big] \nnb \\
\ar 4 m_\ell (1 - r - \hat s) \mbox{\rm Re}\Big[{\cal D}_{1}^\ast {\cal S}_{3}+{\cal E}_{1}^\ast
 {\cal K}_{3}+{\cal D}_{3}^\ast {\cal S}_{1}+{\cal E}_{3}^\ast {\cal K}_{1}\Big] \nnb \\
\ar 2 (1 - r - \hat s) (4 m_\ell^2 - m_{\Lambda_b}^2 \hat s) \, \mbox{\rm Re}\Big[{\cal G}_{1}^\ast
 {\cal H}_{3}+{\cal H}_{1}^\ast {\cal G}_{3}\Big] \nnb \\
\ek m_{\Lambda_b} \Big[(1-r)^2 -\hat s(1+r) \Big] \Big( \vel {\cal K}_{1} \ver^2 +  \vel {\cal S}_{1}
 \ver^2 \Big) \Bigg\} \nnb \\
\ek 32 m_{\Lambda_b}^4 \sqrt{r} \hat s \Bigg\{ 2 m_\ell \mbox{\rm Re}\Big[{\cal D}_{1}^\ast
 {\cal S}_{1}+{\cal E}_{1}^\ast {\cal K}_{1}\Big]+(4 m_\ell^2 - m_{\Lambda_b}^2 \hat s) \,
\mbox{\rm Re}\Big[{\cal G}_{1}^\ast {\cal H}_{1}\Big] \Bigg\} \nnb \\
\ar 8 m_{\Lambda_b}^6 \hat s^2 \Bigg\{ 4 \sqrt{r} \,\mbox{\rm Re}\Big[{\cal K}_{1}^\ast
 {\cal S}_{1}\Big]+2 m_{\Lambda_b} \sqrt{r} (1 - r + \hat s) \mbox{\rm Re}\Big[{\cal K}_{1}^\ast
 {\cal K}_{3}+{\cal S}_{1}^\ast {\cal S}_{3}\Big] \nnb \\
&+& 2 m_{\Lambda_b} (1 - r - \hat s) \mbox{\rm Re}\Big[{\cal K}_{1}^\ast {\cal S}_{3}
+{\cal S}_{1}^\ast {\cal K}_{3}\Big] \nnb \\
\ek (4 m_\ell^2 - m_{\Lambda_b}^2 \hat s) (1 + r - \hat s) \Big( \vel {\cal G}_{3} \ver^2 + 
 \vel {\cal H}_{3} \ver^2 \Big) \nnb \\
\ek 4 m_\ell (1 + r - \hat s) \mbox{\rm Re}\Big[{\cal D}_{3}^\ast {\cal K}_{3}+{\cal E}_{3}^\ast
 {\cal S}_{3}\Big]- 8 m_\ell \sqrt{r} \mbox{\rm Re}\Big[{\cal D}_{3}^\ast {\cal S}_{3}+{\cal E}_{3}^
\ast {\cal K}_{3}\Big]\Bigg\} \nnb \\
\ar 8 m_{\Lambda_b}^8 \hat s^3 \Bigg\{ (1 + r - \hat s) \Big( \vel {\cal K}_{3} \ver^2 +  
\vel {\cal S}_{3} \ver^2 \Big) + 4 \sqrt{r} \mbox{\rm Re}\Big[{\cal K}_{3}^\ast {\cal S}_{3}
\Big]\Bigg\}, \nnb \\
\end{eqnarray}
\begin{eqnarray} {\cal T}_{1}(\hat s) &=& -32
m_{\lb}^4 m_\ell \sqrt{\lambda} v (1 - r)
Re\Big({\cal A}_{1}^* {\cal G}_{1}+{\cal B}_{1}^* {\cal H}_{1}\Big)\nn\\
&-&16 m_{\lb}^4\s1 v \sqrt{\lambda} 
\Bigg\{ 2 Re\Big({\cal A}_{1}^* {\cal D}_{1}\Big)-2Re\Big({\cal B}_{1}^* {\cal E}_{1}\Big)\nn\\
&+&2 m_{\lb} Re\Big({\cal B}_{1}^* {\cal D}_{2}-{\cal B}_{2}^* {\cal D}_{1}+{\cal A}_{2}^* 
{\cal E}_{1}-{\cal A}_{1}^*{\cal E}_{2}\Big)\nn\\
&+&2 m_{\lb} m_\ell Re\Big({\cal A}_{1}^* {\cal H}_{3}+{\cal B}_{1}^* {\cal G}_{3}-{\cal A}_{2}^*
 {\cal H}_{1}-{\cal B}_{2}^*{\cal G}_{1}\Big)\Bigg\}\nn\\
&+&32 m_{\lb}^5 \s1~ v \sqrt{\lambda} \Bigg\{
m_{\lb} (1-r)Re\Big({\cal A}_{2}^* {\cal D}_{2} -{\cal B}_{2}^* {\cal E}_{2}\Big)\nn\\
&+& \sqrt{r} Re\Big({\cal A}_{2}^* {\cal D}_{1}+{\cal A}_{1}^* {\cal D}_{2}-
{\cal B}_{2}^*{\cal E}_{1}-{\cal B}_{1}^* {\cal E}_{2}\Big)\nn\\
&-& \sqrt{r} m_\ell Re\Big({\cal A}_{1}^* {\cal G}_{3}+{\cal B}_{1}^* {\cal H}_{3}+
{\cal A}_{2}^*{\cal G}_{1}+{\cal B}_{2}^* {\cal H}_{1}\Big)\Bigg\} \nn\\
&+&32 m_{\lb}^6 m_\ell \sqrt{\lambda} v \hat s^2 Re\Big({\cal A}_{2}^* {\cal G}_{3}+
{\cal B}_{2}^* {\cal H}_{3}\Big),\nn\\
\end{eqnarray}\
and
\begin{eqnarray} {\cal T}_{2}(\hat s) \es - 8 m_{\Lambda_b}^4 v^2 \lambda \Big(\vel {\cal A}_{1}
 \ver^2 + \vel {\cal B}_{1} \ver^2 + \vel {\cal D}_{1} \ver^2 + \vel {\cal E}_{1} \ver^2 \Big) 
\nnb \\
\ar 8 m_{\Lambda_b}^6 \hat s v^2 \lambda \Big( \vel {\cal A}_{2} \ver^2 + \vel
{\cal B}_{2} \ver^2 + \vel {\cal D}_{2} \ver^2 + \vel {\cal E}_{2} \ver^2 \Big) ~.\nn\\ 
\end{eqnarray}\

We integrate Eq.\eqref{DDR} over $z$ in the interval $[-1,1]$ in order to obtain the
 differential decay width only with respect to $\hat s$. Consequently, we get
 \begin{eqnarray} \label{Decaywitdh}
\frac{d\Gamma}{d \hat s} (\hat s)= \frac{G_F^2\alpha^2_{em} m_{\Lambda_b}}{8192
\pi^5}| V_{tb}V_{ts}^*|^2 v \sqrt{\lambda} \, \Bigg[{{\cal T}_0(\hat s)
+\frac{1}{3} {\cal T}_2(\hat s)}\Bigg]~. \label{decayrate} \end{eqnarray}
\subsection{The differential branching ratio}
In this part, we would like to analyze the differential branching ratio of the transition under 
consideration at different lepton channels. For this aim, using the differential decay width in Eq.\eqref{Decaywitdh}, 
we discuss the variation of the differential branching ratio with respect to $q^{2}$ and other related parameters.
For this aim, we need some the input parameters as presented in Tables 1 and 2.
\begin{table}[ht!]
\centering
\rowcolors{1}{lightgray}{white}
\begin{tabular}{cc}
\hline \hline
   Input Parameters  &  Values    
           \\
\hline \hline
$ m_{e}                   $          &  $ 0.51\times10^{-3}$ $GeV$  \\
$ m_{\mu}                 $          &  $ 0.1056 $ $GeV$             \\
$ m_{\tau}                $          &  $ 1.776  $ $GeV$              \\
$ m_{c}   $          &  $ 1.275 $  $GeV$               \\
$ m_{b}   $          &  $ 4.18      $ $GeV$             \\
$ m_{t}  $          &  $ 160      $ $GeV$               \\
$ m_{W}                   $          &  $ 80.4     $ $GeV$                \\
$ m_{Z}                   $          &  $ 91.2     $ $GeV$                 \\
$ m_{\Lambda_b}           $          &  $ 5.620    $ $GeV$                  \\
$ m_{\Lambda}             $          &  $ 1.1156   $ $GeV$                   \\
$ \tau_{\Lambda_b}        $          &  $ 1.425\times 10^{-12} $ $s$          \\
$ \hbar                   $          &  $ 6.582\times 10^{-25} GeV s $         \\
$ G_{F}                   $          &  $ 1.17\times 10^{-5} $ $GeV^{-2}$       \\
$ \alpha_{em}             $          &  $ 1/137 $                                \\
$ | V_{tb}V_{ts}^*|       $          &  $ 0.041 $                                 \\
 \hline \hline
\end{tabular}
\caption{The values of some input parameters ​​used in the numerical analysis. The quarks masses are in $\overline{MS}$
 scheme \cite{PDG}.}
\end{table}

\begin{table}[ht!]
\centering
\rowcolors{1}{lightgray}{white}
\begin{tabular}{cc}
\hline \hline
   Input Parameters  &  Values    
           \\
\hline \hline
$ m_{\pi_{t}^{0}}   $          &  $(200-500)  $ $GeV$      \\
$ m_{\pi_{t}^{+}}   $          &  $(350-600)  $ $GeV$       \\
$ m_{h_{t}^{0}}     $          &  $(200-500)  $ $GeV$        \\
$ M_{Z^{\prime}}    $          &  $(1200-1800)$ $GeV$         \\
$ F_{\pi}           $          &  $ 50        $ $GeV$          \\
$ \varepsilon       $          &  $(0.06-0.1) $                 \\
$ K_{1}             $          &  $(0.3-1)    $                  \\
 \hline \hline
\end{tabular}
\caption{The values of some input parameters related to TC2 model ​​used in the numerical analysis \cite{HeffTC2}.}
\end{table}
We also use the typical values $ m_{\pi_{t}^{0}}=m_{h_{t}^{0}}=300~GeV$, 
$ m_{\pi_{t}^{+}}=450~GeV$, $ M_{Z^{\prime}}=1500~GeV$, $ \varepsilon=0.08$ and $ K_{1}=0.4$
in our numerical calculations \cite{HeffTC2}.    

As other input parameters, we need the form factors calculated via light cone 
QCD sum rules in full theory \cite{form-factors}. The fit function for 
$f_{1}$, $f_{2}$, $f_{3}$, $g_{1}$, $g_{2}$, $g_{3}$, $f^T_{2}$, $f^T_{3}$, $g^T_{2}$ and $g^T_{3}$ 
is given by \cite{form-factors} (for an alternative parametrization of form factors see \cite{Feldmann:2011xf})
\begin{eqnarray}\label{formfactors}
 f^{(T)}_{i}(q^2)[g^{(T)}_{i}(q^2)]=\frac{a}{\Bigg(1-\ds\frac{q^2}{m_{fit}^2}\Bigg)}+
\frac{b}{\Bigg(1-\ds\frac{q^2}{m_{fit}^2}\Bigg)^2}~,
\end{eqnarray}
where the $a,~b$ and $m_{fit}^2$ are the fit parameters presented in Table 3. The values of corresponding
form factors at $q^2=0$ are also presented in Table 3.
\begin{table}[ht!]
\centering
\rowcolors{1}{lightgray}{white}
\begin{tabular}{ccccc}
\hline \hline
                & $\mbox{a} $ & $ \mbox{b} $ & $m_{fit}^2$ & \mbox{form factors at} $q^2=0$  \\
\hline \hline
 $ f_1    $        & $ -0.046 $ & $   0.368  $  & $ 39.10 $ & $ 0.322 \pm 0.112$ \\
 $ f_2     $       & $  0.0046 $ & $ -0.017 $  & $  26.37 $ & $-0.011 \pm 0.004 $  \\
 $ f_3      $      & $  0.006 $ & $  -0.021 $  & $  22.99 $ & $-0.015 \pm 0.005 $ \\
 $ g_1       $     & $ -0.220 $ & $   0.538 $  & $  48.70 $ & $0.318 \pm 0.110 $\\
 $ g_2       $     & $  0.005  $ & $ -0.018 $  & $  26.93 $ & $-0.013 \pm 0.004$   \\
 $ g_3        $    & $  0.035  $ & $ -0.050 $ &   $ 24.26 $ & $-0.014 \pm 0.005$ \\
 $ f_2^{T}    $    & $ -0.131 $ & $   0.426 $   & $ 45.70 $ & $0.295 \pm 0.105 $ \\
 $ f_3^{T}    $    & $ -0.046  $ &  $ 0.102 $   & $ 28.31 $ & $0.056 \pm 0.018 $ \\
 $ g_2^{T}     $   & $ -0.369 $  & $  0.664 $   & $ 59.37 $  & $0.294 \pm 0.105$ \\
 $ g_3^{T}     $   & $ -0.026 $ & $  -0.075 $   & $ 23.73 $ & $-0.101 \pm 0.035$ \\
\hline \hline
\end{tabular}
\caption{The fit parameters $a$, $b$ and $m_{fit}^2$ appear in the fit function of the form factors 
$f_{1}$, $f_{2}$, $f_{3}$, $g_{1}$, $g_{2}$, $g_{3}$, $f^T_{2}$, $f^T_{3}$, $g^T_{2}$ and $g^T_{3}$ 
together with the values of the corresponding form factors at $q^2=0$  in full theory for  $\Lambda_b
\rightarrow \Lambda \ell^{+} \ell^{-}$ decay \cite{form-factors}.}
\end{table}
In addition, the fit function of the form factors $f^T_{1}$ and $g^T_{1}$ is given by
 \cite{form-factors}
\begin{eqnarray}\label{frmfctft1gt1}
 f^T_{1}(q^2)[g^T_{1}(q^2)]=\frac{c}{\Bigg(1-\ds\frac{q^2}{m_{fit}^{' 2}}\Bigg)}-
\frac{c}{\Bigg(1-\ds\frac{q^2}{m_{fit}^{'' 2}}\Bigg)^2}~, 
\end{eqnarray}
where the $c$, $m^{'2}_{fit}$ and $m^{''2}_{fit}$ are the fit parameters which we present their values
together with the values of corresponding form factors at $q^2=0$ in Table 4.
\begin{table}[ht!]
\centering
\rowcolors{1}{lightgray}{white}
\begin{tabular}{ccccc}
\hline \hline
           & $\mbox{c} $ & $ m_{fit}^{' 2}$ & $ m_{fit}^{'' 2} $ & \mbox{form factors at} $q^2=0$ \\
          \hline \hline
$ f_1^{T} $ &   $ -1.191 $  & $ 23.81 $   & $ 59.96  $ &  $0     \pm 0.0$      \\
$ g_1^{T} $ &   $ -0.653 $ & $ 24.15 $  & $ 48.52   $  & $0     \pm 0.0$ \\
 \hline \hline
 \end{tabular}
\caption{The fit parameters $c$, $ m_{fit}^{' 2}$ and $ m_{fit}^{'' 2} $ in the fit function of the form factors
$f^T_{1}$ and  $g^T_{1}$ together with the values of the related form factors at $q^2=0$  in full theory for  $\Lambda_b
\rightarrow \Lambda \ell^{+} \ell^{-}$ decay \cite{form-factors}.}
\end{table}

The dependences of the differential branching ratio on $q^{2}$, $m_{\pi_{t}^{+}}$ and 
$M_{Z^{\prime}}$ in the cases of $\mu$ and $\tau$ leptons in both SM and TC2 models are shown in Figures 1-6. 
In each figure we show the dependences of the differential branching ratio on different observables
for both central values of the form factors (left panel) and form factors with their uncertainties (right panel).
Note that, the results for the case of $e$ are very close to those of $\mu$, so we do not present the results 
for $e$ in our figures. We also depict the recent experimental results on the differential branching ratio in
$\mu$ channel provided by the CDF \cite{CDF} and LHCb \cite{LHCb} Collaborations in Figure 1.
 From these figures, we conclude that
 \begin{itemize}
\item for both lepton channels, there are considerable differences between predictions of the SM and
 TC2 models on the differential branching ratio with respect to $q^{2}$, $m_{\pi_{t}^{+}}$ and
 $M_{Z^{\prime}}$ when the central values of the form factors are considered. 
 \item Although the swept regions by both models coincide somewhere, but adding the uncertainties of form factors
 can not totally  kill the differences between two models predictions on the differential branching ratio. 
 \item In the case of differential branching ratio in terms of $q^2$ at $\mu$ channel (Figure 1), the experimental
 data by CDF and LHCb Collaborations are overall close to the SM predictions for $q^2 \leq 16 ~GeV^2$. When 
 $q^2 > 16 ~GeV^2$ the experimental data lie in the common region swept by the SM and TC2 models.
\end{itemize}
\begin{figure}[ht!]
\centering
\begin{tabular}{ccc}
\epsfig{file=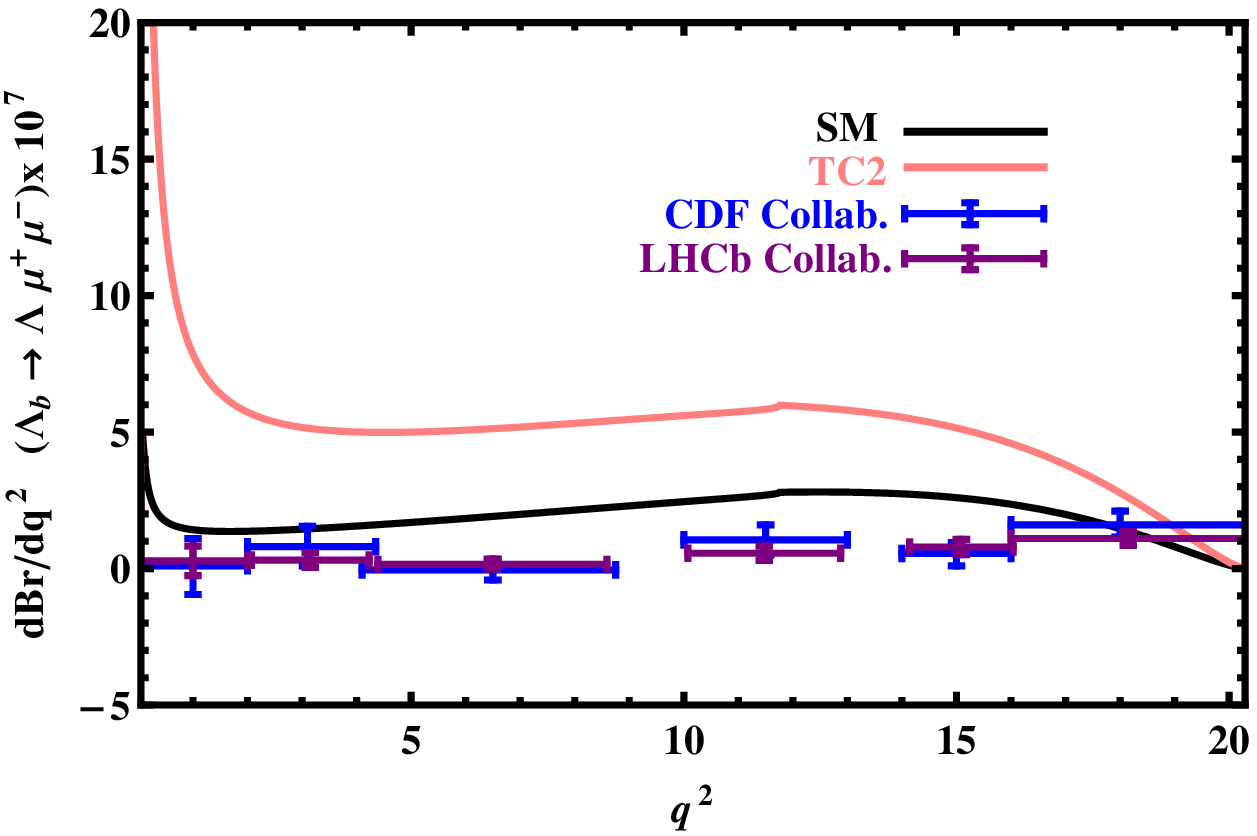,width=0.45\linewidth,clip=}&
\epsfig{file=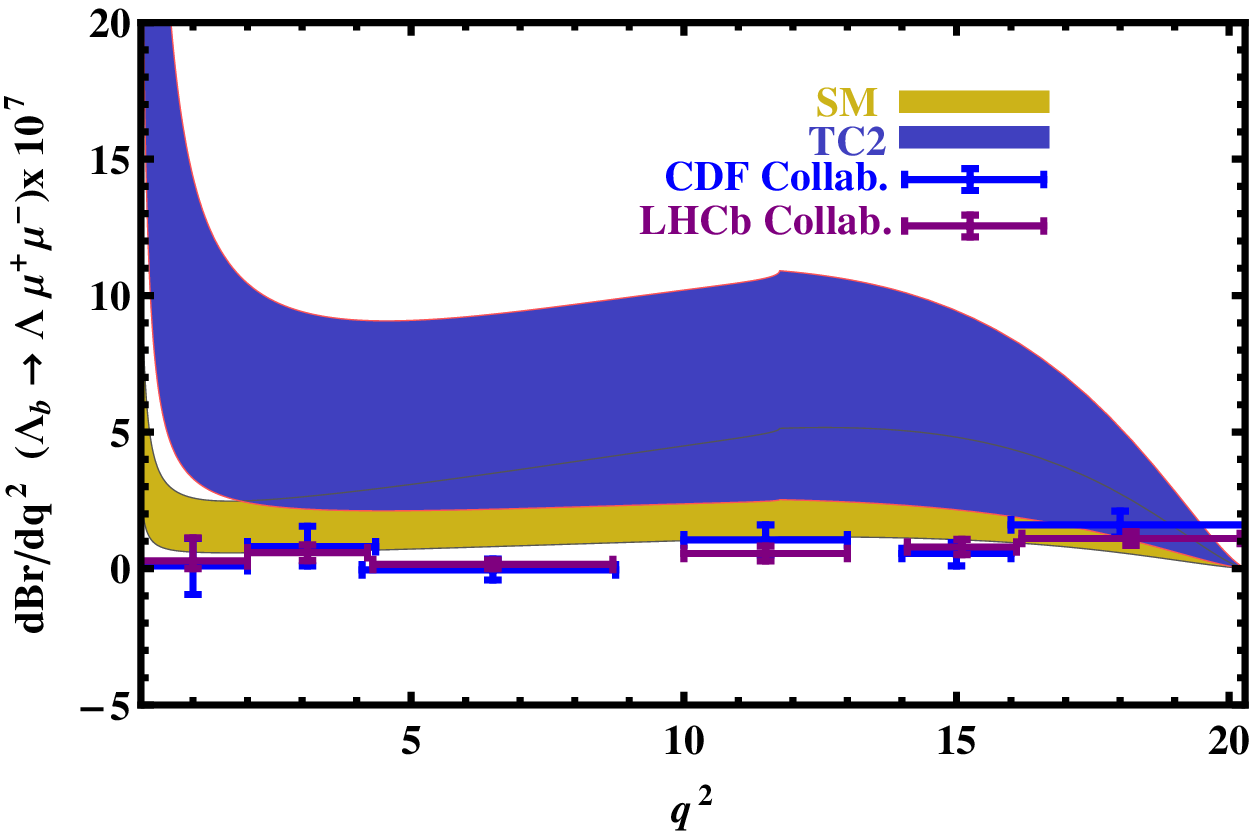,width=0.45\linewidth,clip=}
\end{tabular}
\caption{The dependence of the differential branching ratio (in $GeV^{-2}$ unit) on $q^{2}$ ($GeV^2$) for $\Lambda_b
\rightarrow \Lambda \mu^{+} \mu^{-}$ decay channel in the SM and TC2 models using the central values 
of the form factors (left panel) and form factors with their uncertainties (right panel). The recent 
experimental results by CDF \cite{CDF} and LHCb \cite{LHCb} are also presented in both figures.}
\end{figure}
\begin{figure}[ht!]
\centering
\begin{tabular}{ccc}
\epsfig{file=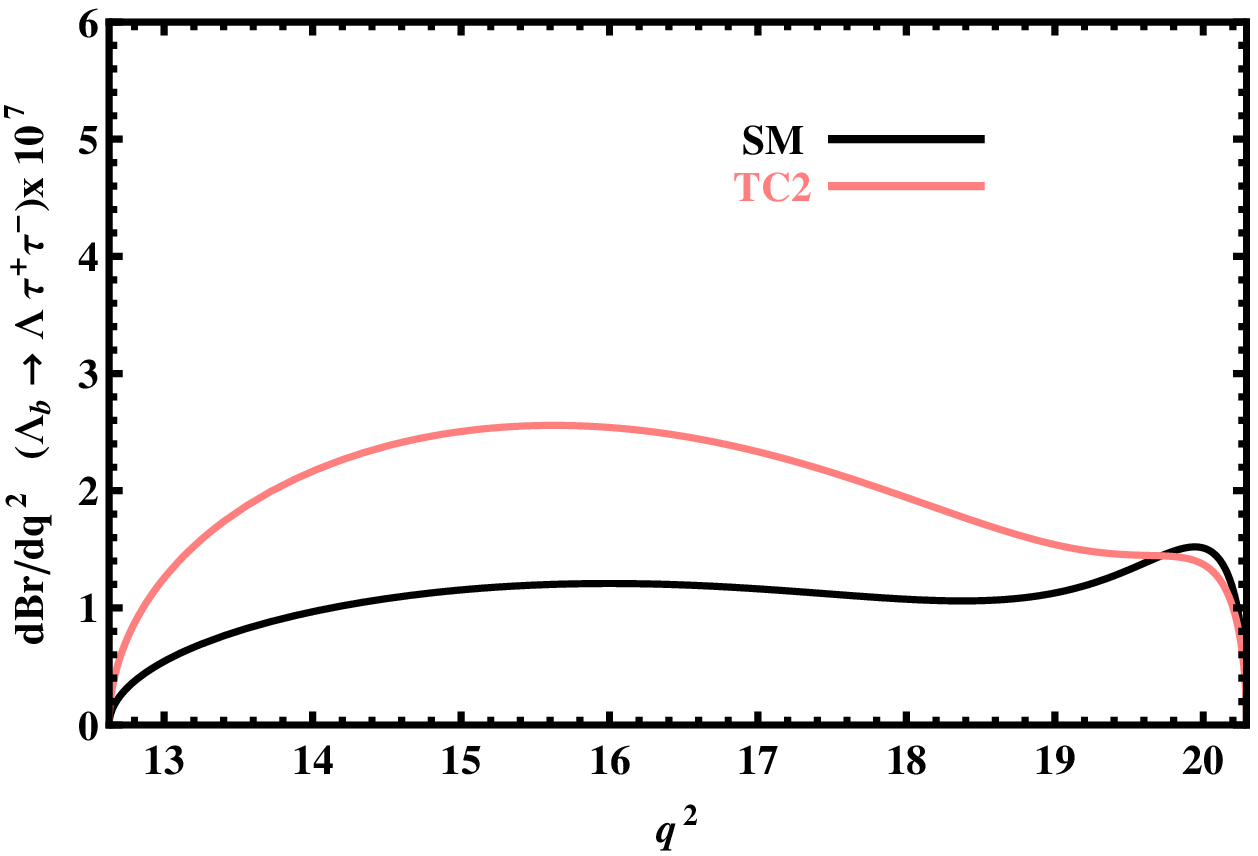,width=0.45\linewidth,clip=}&
\epsfig{file=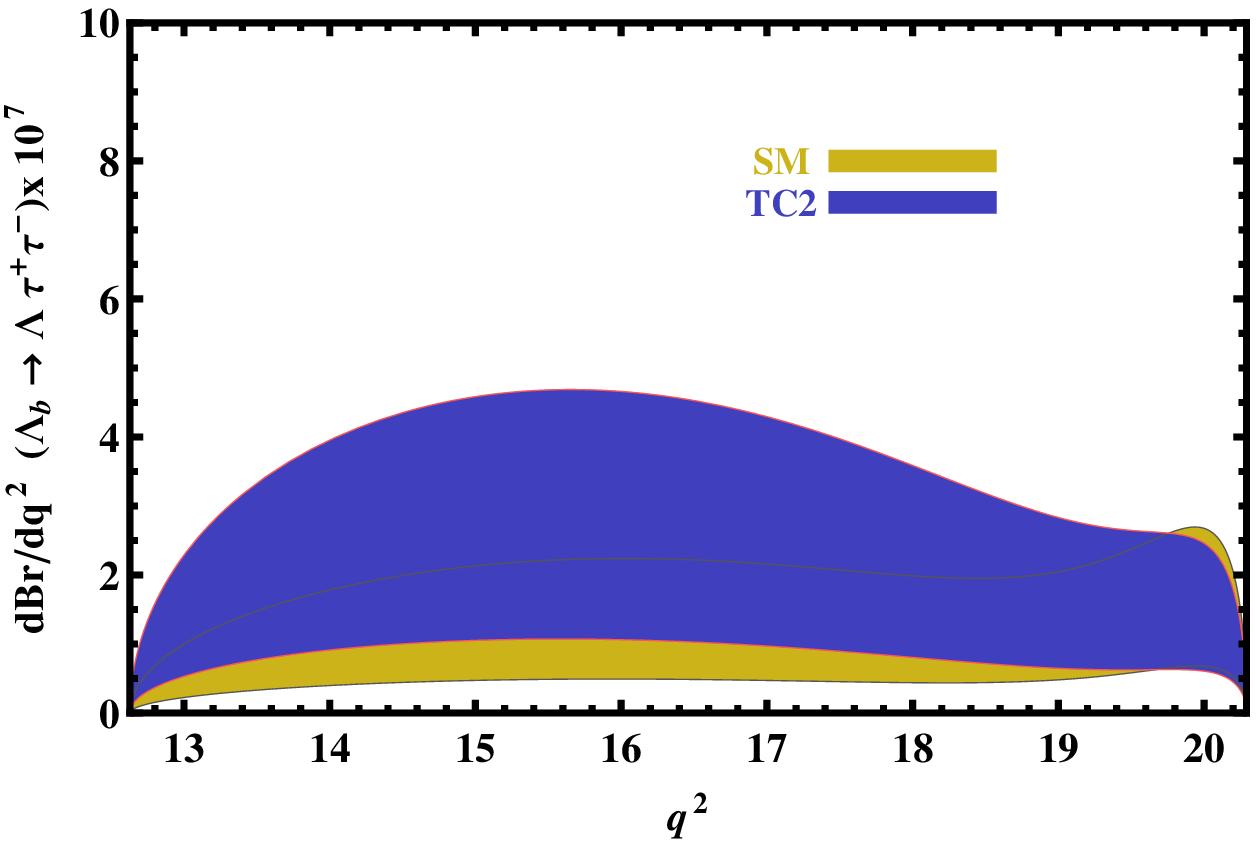,width=0.45\linewidth,clip=}
\end{tabular}
\caption{The dependence of the differential branching ratio (in $GeV^{-2}$ unit) on $q^{2}$ ($GeV^2$) for $\Lambda_b
\rightarrow \Lambda \tau^{+} \tau^{-}$ decay channel in the SM and TC2 models using the central values 
of the form factors (left panel) and form factors with their uncertainties (right panel).}
\end{figure}
\begin{figure}[ht!]
\centering
\begin{tabular}{ccc}
\epsfig{file=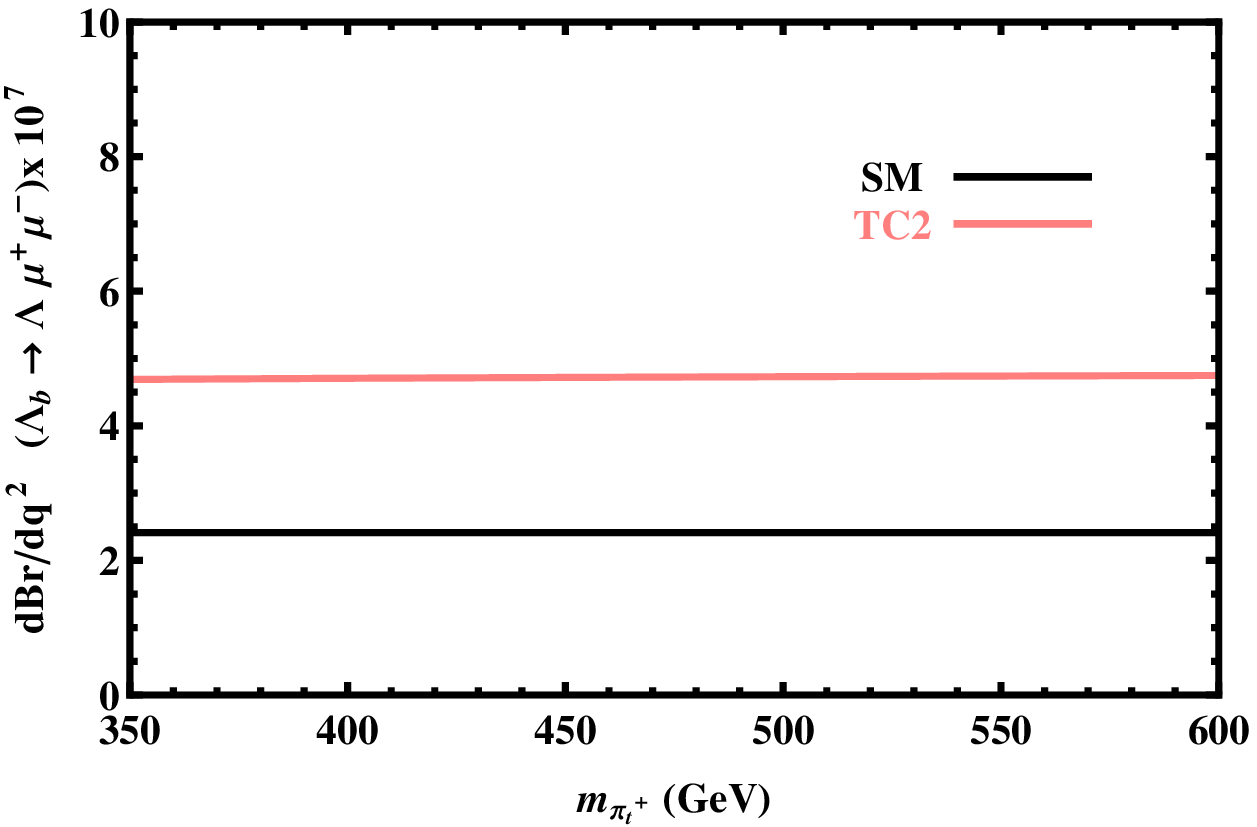,width=0.45\linewidth,clip=}&
\epsfig{file=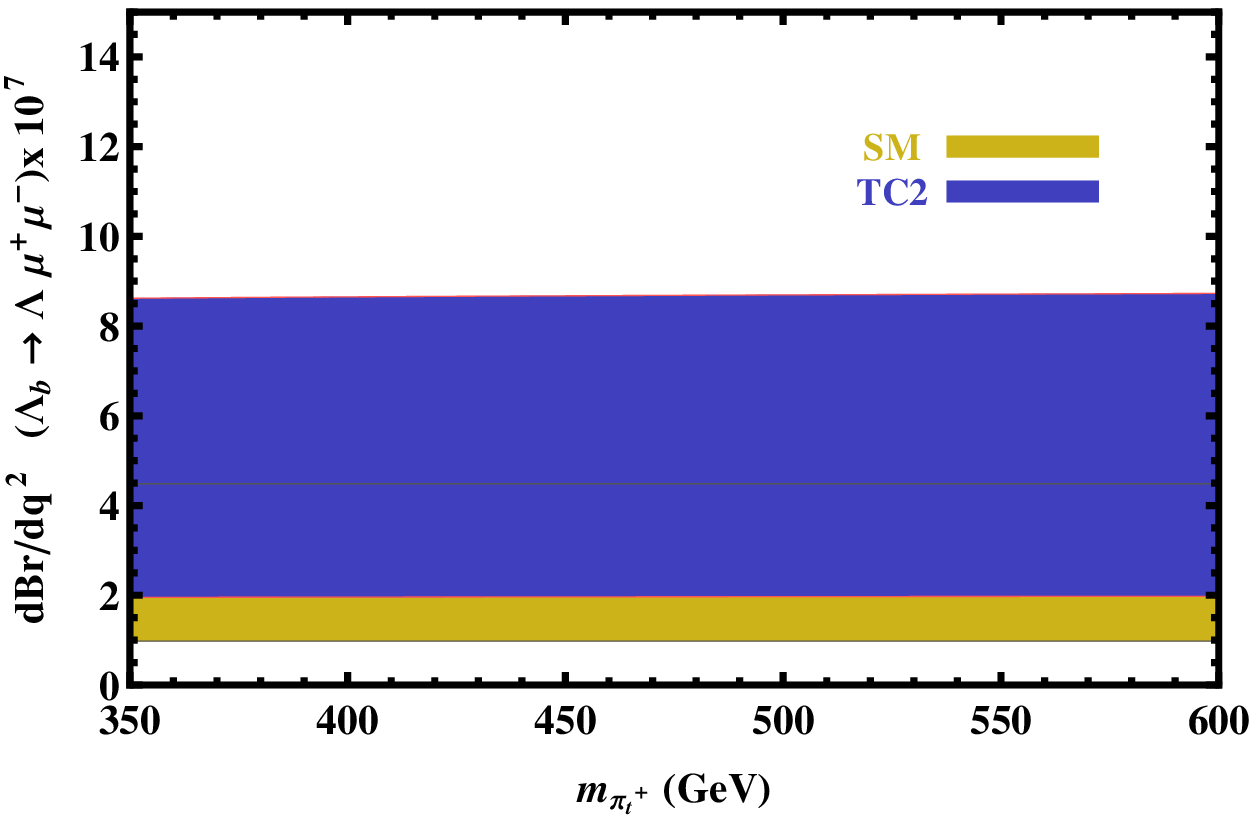,width=0.45\linewidth,clip=}
\end{tabular}
\caption{The dependence of the differential branching ratio (in $GeV^{-2}$ unit) on $m_{\pi_{t}^{+}}$ for
 $\Lambda_b\rightarrow \Lambda \mu^{+} \mu^{-}$ decay channel in the SM and TC2 models using the central values 
of the form factors (left panel) and form factors with their uncertainties (right panel).}
\end{figure}
\begin{figure}[ht!]
\centering
\begin{tabular}{ccc}
\epsfig{file=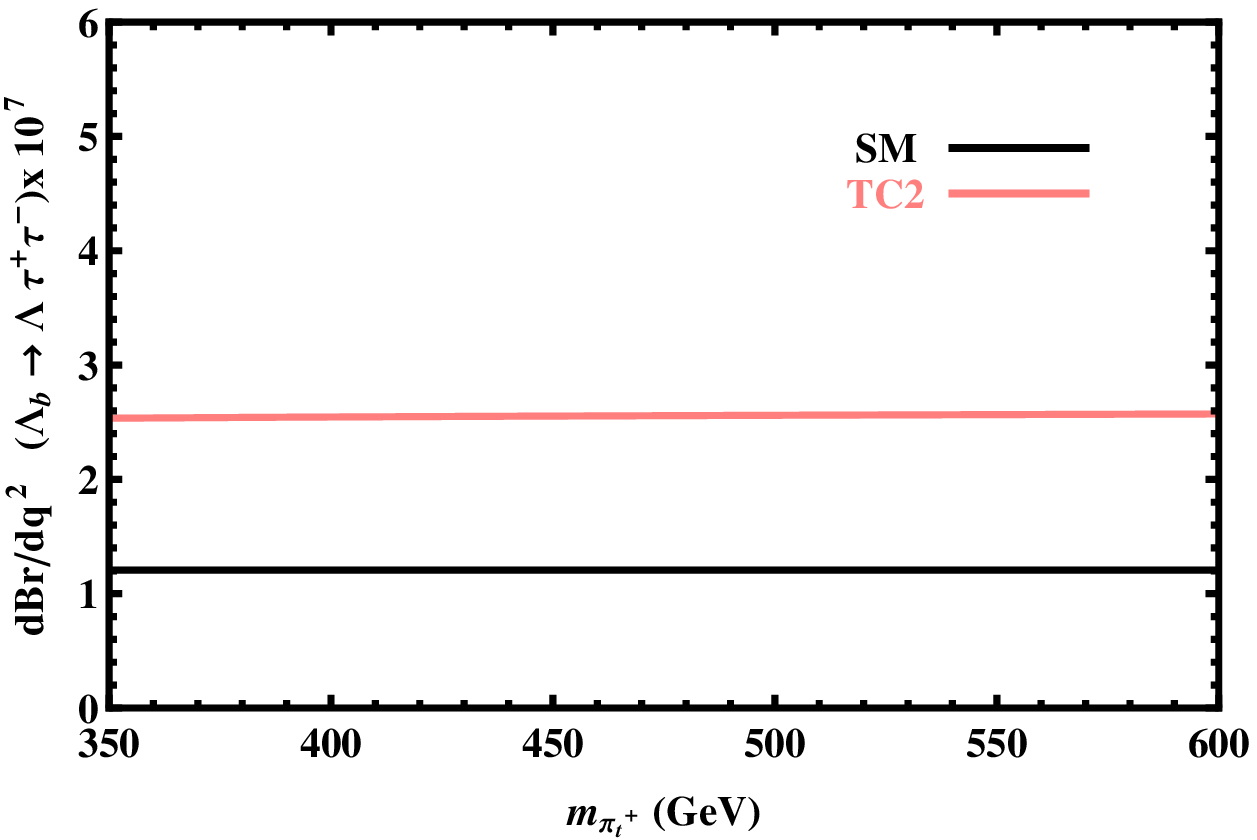,width=0.45\linewidth,clip=}&
\epsfig{file=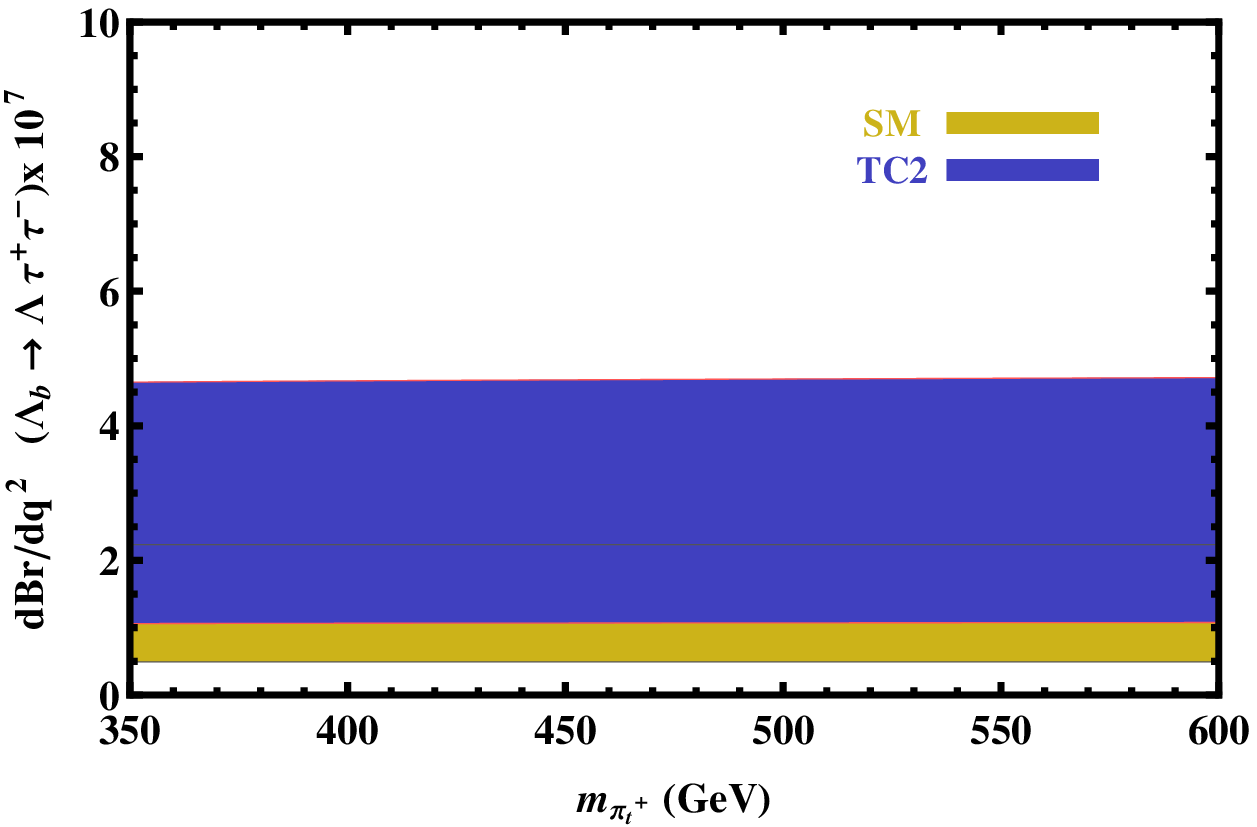,width=0.45\linewidth,clip=}
\end{tabular}
\caption{The same as figure 3 but for $\Lambda_b
\rightarrow \Lambda \tau^{+} \tau^{-}$ decay channel.}
\end{figure}
\begin{figure}[ht!]
\centering
\begin{tabular}{cc}
\epsfig{file=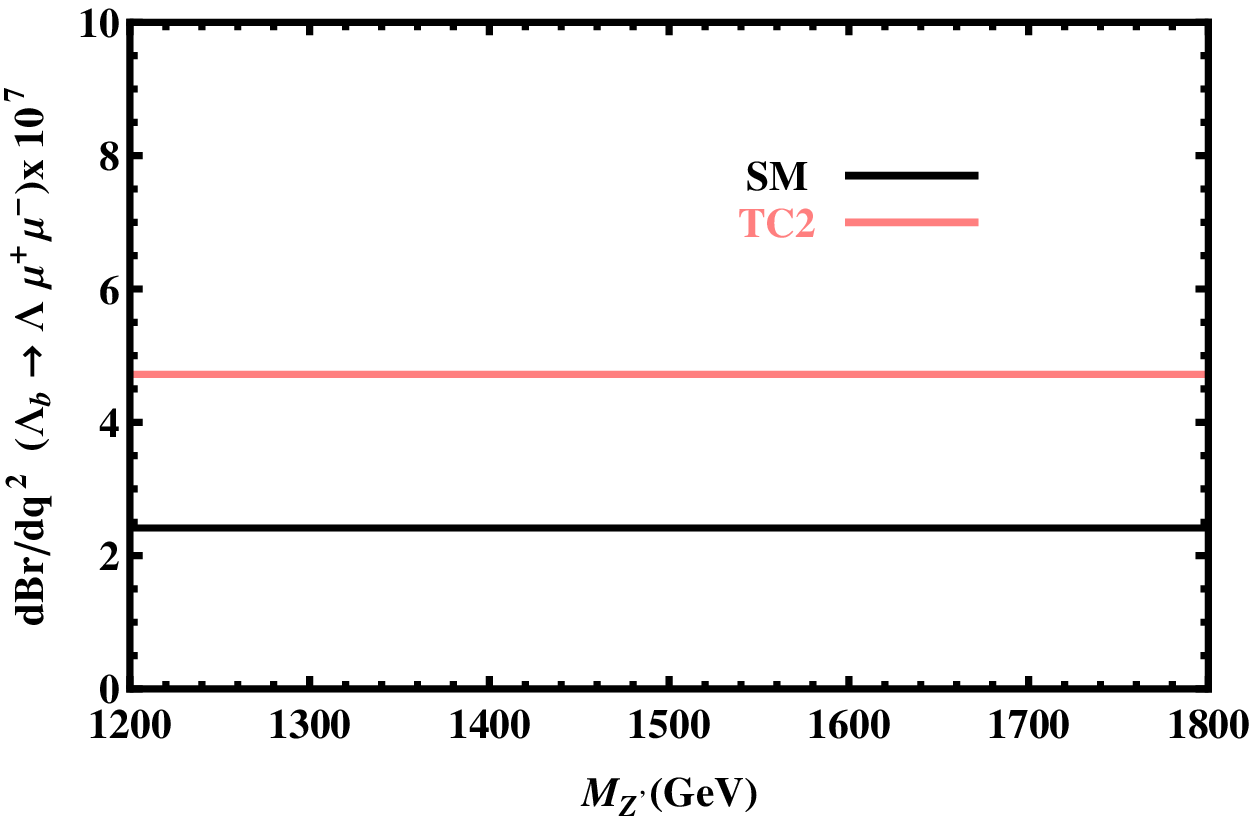,width=0.45\linewidth,clip=}&
\epsfig{file=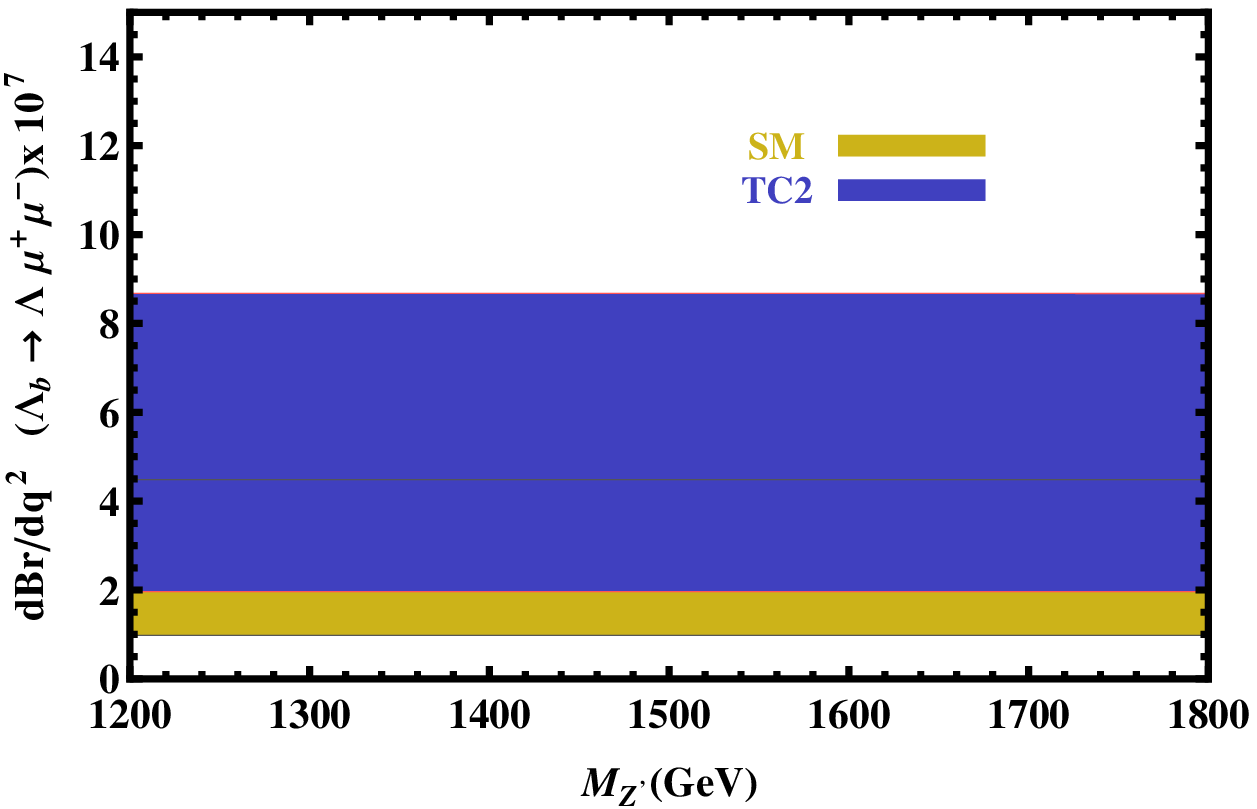,width=0.45\linewidth,clip=}
\end{tabular}
\caption{The dependence of the differential branching ratio (in $GeV^{-2}$ unit) on $M_{Z^{\prime}}$ for
 $\Lambda_b\rightarrow \Lambda \mu^{+} \mu^{-}$ decay channel in the SM and TC2 models using the central values 
of the form factors (left panel) and form factors with their uncertainties (right panel).}
\end{figure}
\begin{figure}[ht!]
\centering
\begin{tabular}{cc}
\epsfig{file=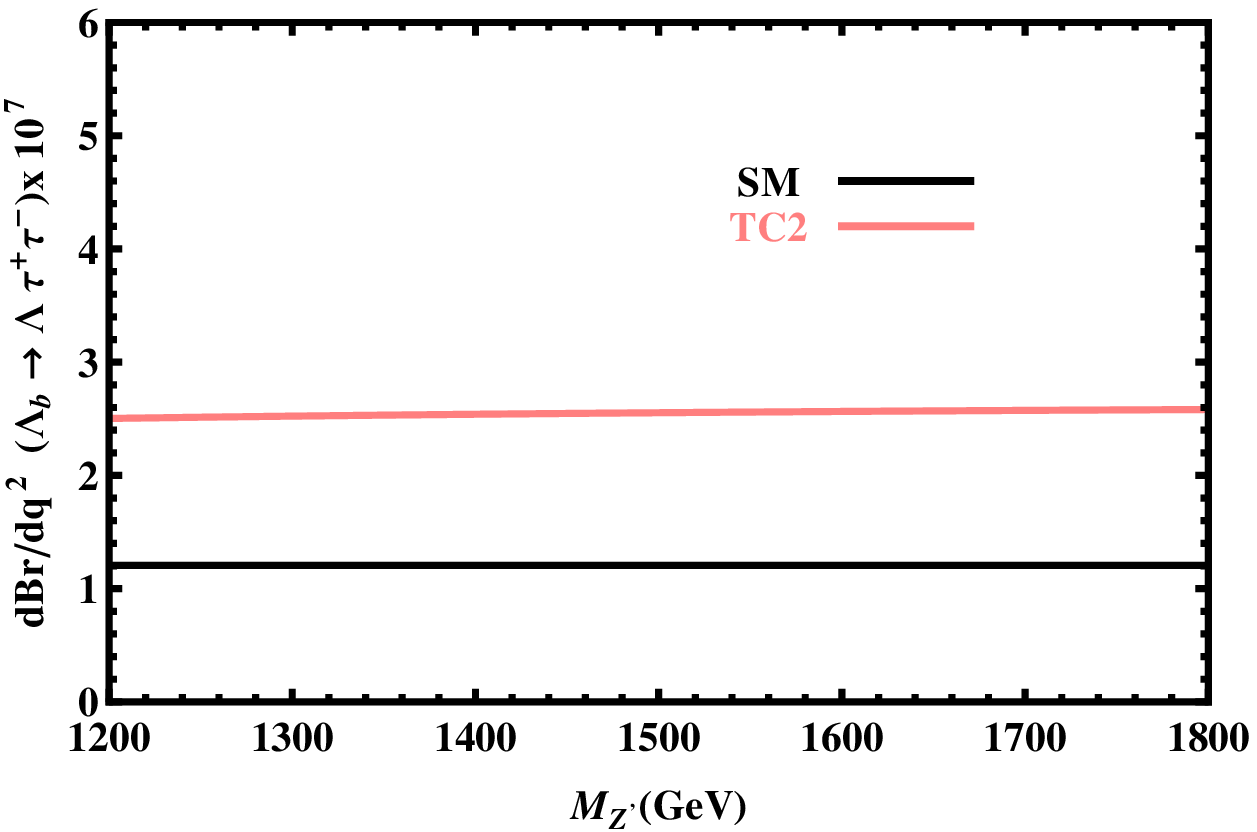,width=0.45\linewidth,clip=}&
\epsfig{file=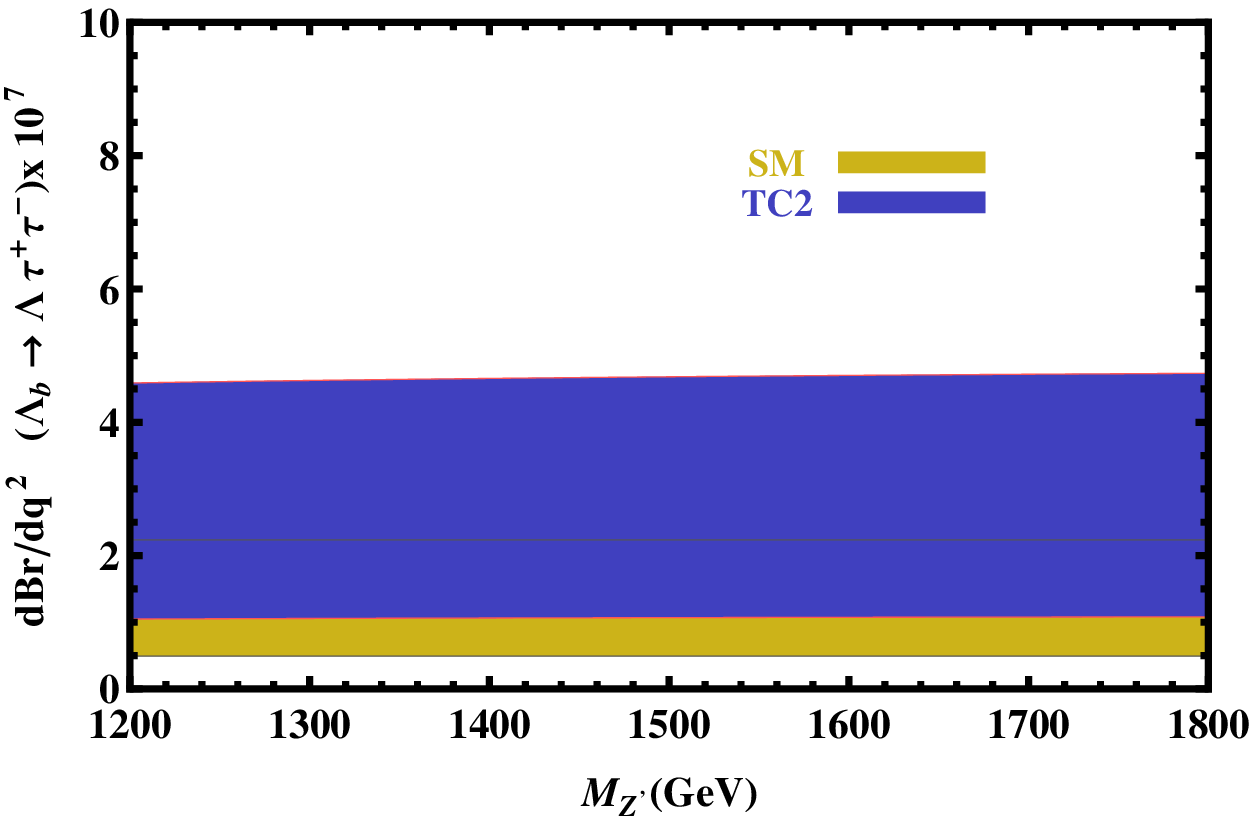,width=0.45\linewidth,clip=}
\end{tabular}
\caption{The same as figure 5 but for $\Lambda_b
\rightarrow \Lambda \tau^{+} \tau^{-}$ decay channel.}
\end{figure}
To better compare the results, we depict the numerical values of the differential  branching ratio  at different values of $q^2$ in its allowed region for all lepton channels and both SM and TC2 models 
in Tables (5-7). We also present the experimental data in $\mu$ channel provided by the CDF \cite{CDF} and LHCb \cite{LHCb} Collaborations in Table 5. With a quick  glance in these tables, we see that
\begin{itemize}
 \item in the case of $\mu$,  the experimental data on the differential branching ratio especially those provided by the CDF Collaboration coincide/are close with/to the intervals predicted by the SM in all ranges of the $q^2$. Within the errors, 
the results of TC2 are consistent with the  data by CDF Collaboration in the intervals $[2.00-4.30], ~[10.09-12.86]$ and $[16.00-20.30]$   for the $q^2$ and  with the  data by LHCb Collaboration only for the interval
$[16.00-20.30]$   for the $q^2$.
\item In all lepton channels and within the errors, the intervals predicted by the TC2 model for the differential branching ratio coincide partly with the intervals predicted by the SM approximately in all 
ranges of the  $q^2$.
\end{itemize}
These results show that although central values of the theoretical results differ considerably with the experimental data, considering the errors of form factors bring the
 intervals predicted by theory in both models close to the experimental data especially in the case of SM.

\begin{table}[ht!]
\centering
\rowcolors{1}{lightgray}{white}
\begin{tabular}{ccccc}
\hline \hline
 $ $ & {\mbox{\footnotesize SM}} &{\mbox{\footnotesize TC2}} & {\mbox{\footnotesize CDF \cite{CDF}}} &{\mbox{\footnotesize LHCb \cite{LHCb}}} \\
\hline \hline
 {\footnotesize $q^2$} & {\footnotesize $dBr/dq^2[10^{-7}]$} 
& {\footnotesize $dBr/dq^2[10^{-7}]$} & {\footnotesize $dBr/dq^2[10^{-7}]$} &{\footnotesize $dBr/dq^2[10^{-7}]$}\\
\hline \hline
 {\footnotesize $0.00-2.00$ }   &  {\footnotesize $(0.60-2.58)$ }  & {\footnotesize $(3.29-14.36)$ } &  
 {\footnotesize $0.15\pm2.01\pm0.05 $ } & {\footnotesize  $0.28\pm0.38\pm0.40\pm0.06 $ }\\
 {\footnotesize $ 2.00-4.30  $ } & {\footnotesize $(0.61-2.65) $ } & {\footnotesize $(2.16-9.33) $ } &  {\footnotesize
 $1.84\pm1.66\pm0.59$ } & {\footnotesize $0.31\pm0.26\pm0.07\pm0.07 $ }\\
 {\footnotesize $ 4.30-8.68 $ }  & {\footnotesize $(0.80-3.48) $ }  &  {\footnotesize $ (2.17-9.31)  $ } &  
 {\footnotesize $-0.20\pm1.64\pm0.08 $}& {\footnotesize $ 0.15\pm0.17\pm0.02\pm0.03 $ }\\
 {\footnotesize $ 10.09-12.86 $ } & {\footnotesize $ (1.11-4.93) $ }  &  {\footnotesize $ \small{(2.46-10.62)} $ } & {\footnotesize  
 $2.97\pm1.47\pm0.95 $} & {\footnotesize $ 0.56\pm0.21\pm0.16\pm0.12 $ }\\
 {\footnotesize $ 14.18-16.00 $ }  & {\footnotesize $(1.05-4.78) $ }  & {\footnotesize $(2.13-9.37) $ } &  
 {\footnotesize $0.96\pm0.73\pm0.31 $} & {\footnotesize $0.79\pm0.24\pm0.15\pm0.17 $ }\\
 {\footnotesize $16.00-20.30 $ }  & {\footnotesize $(0.54-2.57) $ }  & {\footnotesize $(1.06-4.84) $ } &  
 {\footnotesize $6.97\pm1.88\pm2.23 $ } & {\footnotesize $ 1.10\pm0.18\pm0.17\pm0.24 $ }\\
\hline \hline
\end{tabular}
\caption{Numerical values of the differential branching ratio in $GeV^{-2}$ for different intervals of $q^2$ ($GeV^{2}$)  for
$\Lambda_b \rightarrow \Lambda \mu^+ \mu^-$ decay channel in SM and TC2 models 
obtained using the typical  values of the  masses $m_{\pi^{+}_{t}}=450~GeV$ and 
$M_{Z^{\prime}}=1500~GeV$. We also show the experimental values on the differential branching ratio provided by the CDF \cite{CDF} and LHCb \cite{LHCb} Collaborations in this Table.}
\end{table}

\begin{table}[ht!]
\centering
\rowcolors{1}{lightgray}{white}
\begin{tabular}{ccc}
\hline \hline
 $ $ & {\footnotesize SM} &  {\footnotesize TC2} \\
\hline \hline
  {\footnotesize $q^2$} & {\footnotesize $dBr/dq^2[10^{-7}]$} & {\footnotesize $dBr/dq^2[10^{-7}]$} \\
\hline \hline
{\footnotesize $  0.00-2.00  $}  &  {\footnotesize $(0.60-2.59) $ } & {\footnotesize $ (3.29-14.37) $ } \\
 {\footnotesize $  2.00-4.30  $ } & {\footnotesize $ (0.61-2.65) $ } &{\footnotesize $ (2.17-9.34)  $ } \\
 {\footnotesize $  4.30-8.68  $ } & {\footnotesize $ (0.80-3.49) $}  & {\footnotesize $ (2.17-9.32) $ } \\
 {\footnotesize $ 10.09-12.86 $  } & {\footnotesize  $ (1.11-4.94) $ } & {\footnotesize $ (2.46-10.63) $ } \\
 {\footnotesize $ 14.18-16.00 $ } & {\footnotesize  $ (1.05-4.78) $ } & {\footnotesize $ (2.13-9.37) $ } \\
 {\footnotesize $ 16.00-20.30 $ } & {\footnotesize $ (0.54-2.57) $ } & {\footnotesize $ (1.06-4.84) $ } \\
\hline \hline
\end{tabular}
\caption{Numerical values of the differential branching ratio in $GeV^{-2}$ for different intervals of $q^2$ ($GeV^{2}$)  for
$\Lambda_b \rightarrow \Lambda e^+ e^-$ decay channel in SM and TC2 models 
obtained using the typical  values of the  masses $m_{\pi^{+}_{t}}=450~GeV$ and 
$M_{Z^{\prime}}=1500~GeV$.}
\end{table}

\begin{table}[ht!]
\centering
\rowcolors{1}{lightgray}{white}
\begin{tabular}{ccc}
\hline \hline
 $ $ & {\footnotesize SM} &  {\footnotesize TC2} \\
\hline \hline
  {\footnotesize $q^2$} & {\footnotesize $dBr/dq^2[10^{-7}]$}  
& {\footnotesize $dBr/dq^2[10^{-7}]$}  \\
\hline \hline
{\footnotesize $ 12.60-12.86 $}  & {\footnotesize $ (0.11-0.53) $ }  & {\footnotesize $ (0.28-1.22) $ } \\
{\footnotesize $ 14.18-16.00 $ } & {\footnotesize  $ (0.47-2.15) $ } & {\footnotesize $ (1.05-4.61) $ } \\
{\footnotesize $ 16.00-20.30 $ } & {\footnotesize $ (0.43-1.96) $ }  & {\footnotesize $ (0.77-3.46) $ } \\
\hline \hline
\end{tabular}
\caption{Numerical values of the differential branching ratio in $GeV^{-2}$ for different intervals of $q^2$ ($GeV^{2}$)  for
$\Lambda_b \rightarrow \Lambda \tau^+ \tau^-$ decay channel in SM and TC2 models 
obtained using the typical  values of the  masses $m_{\pi^{+}_{t}}=450~GeV$ and 
$M_{Z^{\prime}}=1500~GeV$.}
\end{table}
\subsection{The branching ratio}
In this part, we calculate the values of the branching ratio of the transition under consideration both in  SM
and TC2 models. For this aim, we need to multiply the total decay width by the lifetime of the initial 
baryon $\Lambda_{b}$ and divide by $\hbar$. Taking into account the typical values for $m_{\pi_{t}^{+}}$ 
and $M_{Z^{\prime}}$, we present the numerical results obtained from our calculations for both models together with the existing experimental data provided by the CDF \cite{CDF} and LHCb \cite{LHCb} Collaborations
in Table 8. As can be seen from this Table,
\begin{itemize}
 \item although the central values of the branching ratios in TC2 model are roughly ($2-3.5$ times) bigger than those of the SM, adding the errors of form factors ends up in coinciding the intervals of 
the values predicted by the two models for all lepton channels.
\item The orders of branching ratio show that the $\Lambda_b\rightarrow \Lambda \ell^{+} \ell^{-}$ decay
can be accessible at the LHC for all leptons. As already
 mentioned, this decay at $\mu$ channel has been previously observed by the CDF and LHCb Collaborations.
\item As it is expected, the value of the branching ratio decreases with increasing the mass of 
the final lepton. 
\item In the case of $\mu$ channel, the experimental data on the branching ratio  coincide with the interval predicted by the SM within the errors, 
but these data considerably differ from the interval predicted by the TC2 model. 
\end{itemize}
\begin{table}[ht!]
\centering
\rowcolors{1}{lightgray}{white}
\begin{tabular}{ccccc}
\hline \hline
{\footnotesize } & {\footnotesize $BR~(\Lambda_b
\rightarrow \Lambda ~e^{+}~ e^{-})[10^{-6}]$ } & {\footnotesize $BR~(\Lambda_b
\rightarrow \Lambda ~\mu^{+}~ \mu^{-})[10^{-6}]$ }& 
{\footnotesize $BR~(\Lambda_b
\rightarrow \Lambda ~\tau^{+} ~\tau^{-})[10^{-6}]$ } \\
\hline \hline
{\footnotesize SM } &  {\footnotesize $(1.81-8.06)  $ }  &{\footnotesize $(1.64-7.30)  $ } &{\footnotesize $(0.34-1.51) $} \\
{\footnotesize TC2 } & {\footnotesize  $(6.62-29.03)$ }  & {\footnotesize $(4.55-19.81)$ } &{\footnotesize $(0.63-2.77) $} \\
{\footnotesize CDF \cite{CDF} } &  $-$& {\footnotesize $1.73\pm0.42\pm0.55$ } & $-$ \\
{\footnotesize LHCb \cite{LHCb} } &  $-$ & {\footnotesize $0.96\pm0.16\pm0.13\pm0.21$ }& $-$\\
\hline \hline
\end{tabular}
\caption{Numerical values of the branching ratio of $\Lambda_b
\rightarrow \Lambda ~\ell^{+} ~\ell^{-}$ for $m_{\pi_{t}^{+}}=450~GeV$ and
 $M_{Z^{\prime}}=1500~GeV$  in SM and TC2 models together with the experimental data provided by the CDF \cite{CDF} and LHCb \cite{LHCb} Collaborations.}
\end{table}
In order to see how the TC2 model predictions deviate from those of SM, we plot the variations of branching ratios
on $m_{\pi_{t}^{+}}$ and $M_{Z^{\prime}}$ in Figures 7-10. From these figures we see that
\begin{itemize}
\item there are big differences between the predictions of SM and TC2 models on branching ratio
 with respect to $m_{\pi_{t}^{+}}$ and $M_{Z^{\prime}}$ when the central values of the form factors 
are considered. 
\item The branching ratios remain approximately unchanged when the masses of $m_{\pi_{t}^{+}}$ and $M_{Z^{\prime}}$ 
are varied in the regions presented in the figures for both leptons.
\item Adding the uncertainties of form factors ending up in intersections between the swept regions of two
models, but can not totally kill the differences between two models predictions.
\end{itemize}
\begin{figure}[ht!]
\centering
\begin{tabular}{ccc}
\epsfig{file=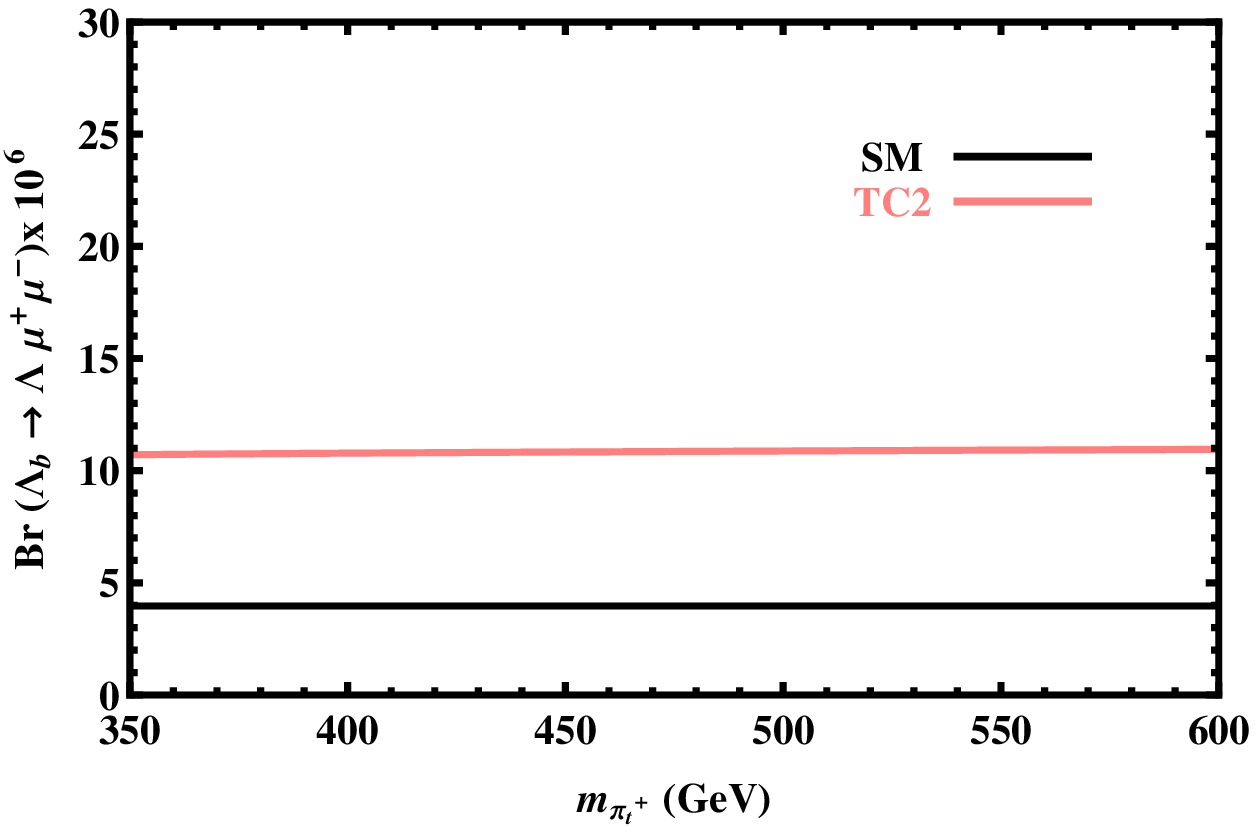,width=0.45\linewidth,clip=}&
\epsfig{file=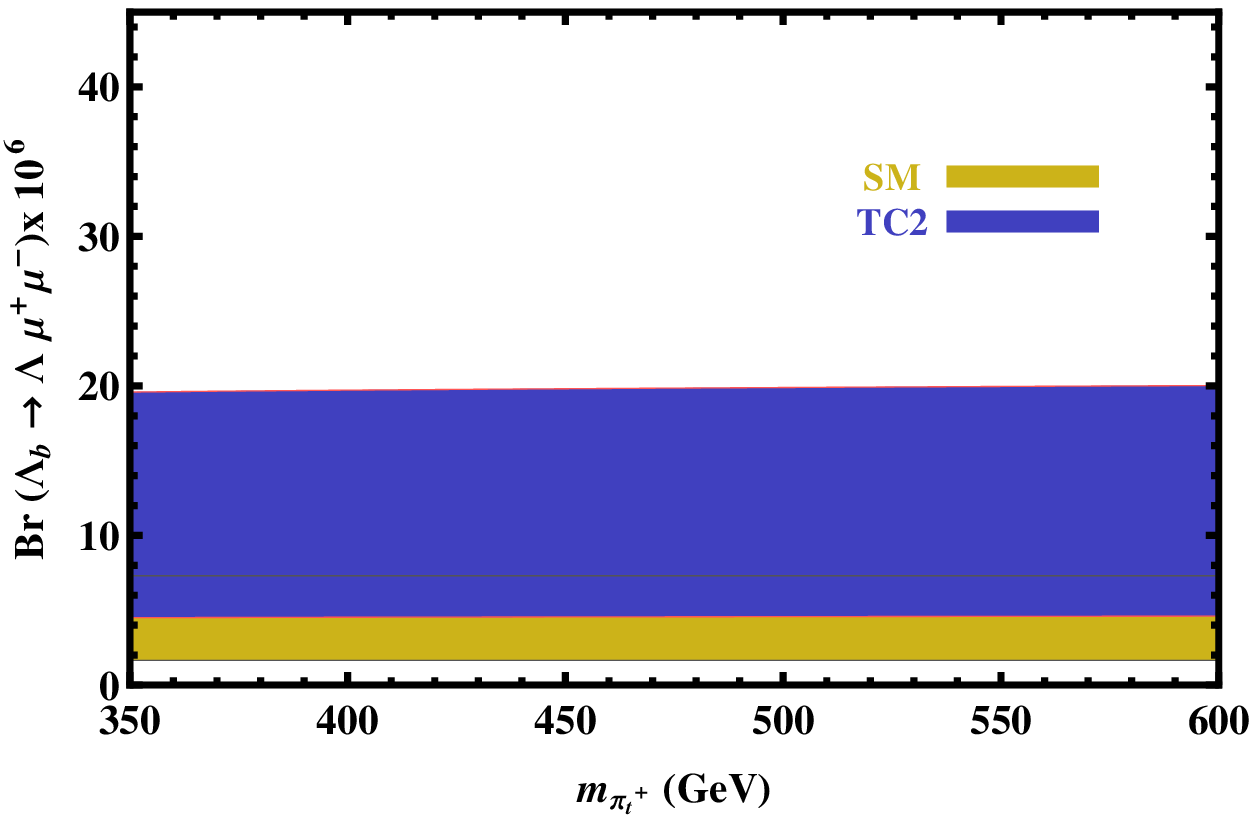,width=0.45\linewidth,clip=}
\end{tabular}
\caption{The dependence of the branching ratio on $m_{\pi_{t}^{+}}$ for
 $\Lambda_b\rightarrow \Lambda \mu^{+} \mu^{-}$ decay channel in the SM and TC2 models using the central 
values of the form factors (left panel) and form factors with their uncertainties (right panel).}
\end{figure}
\begin{figure}[ht!]
\centering
\begin{tabular}{ccc}
\epsfig{file=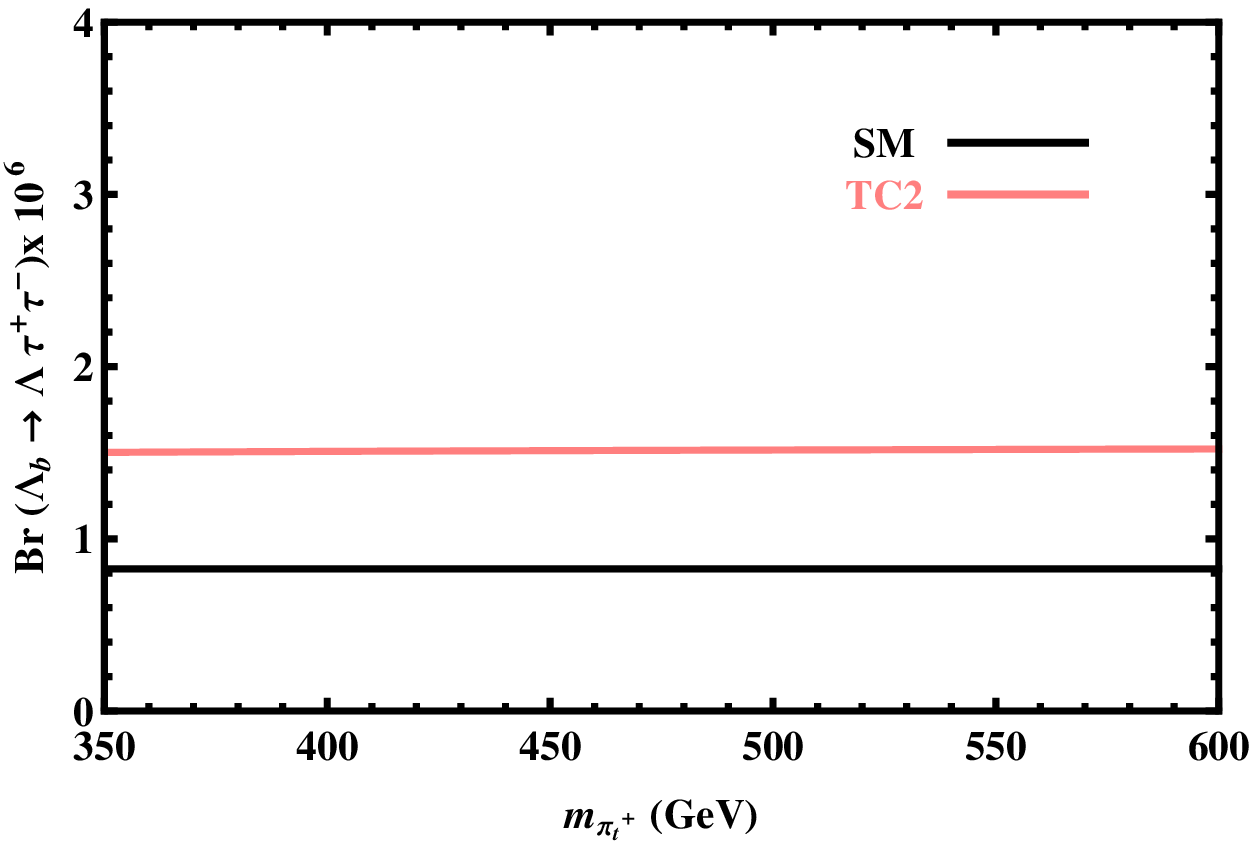,width=0.45\linewidth,clip=}&
\epsfig{file=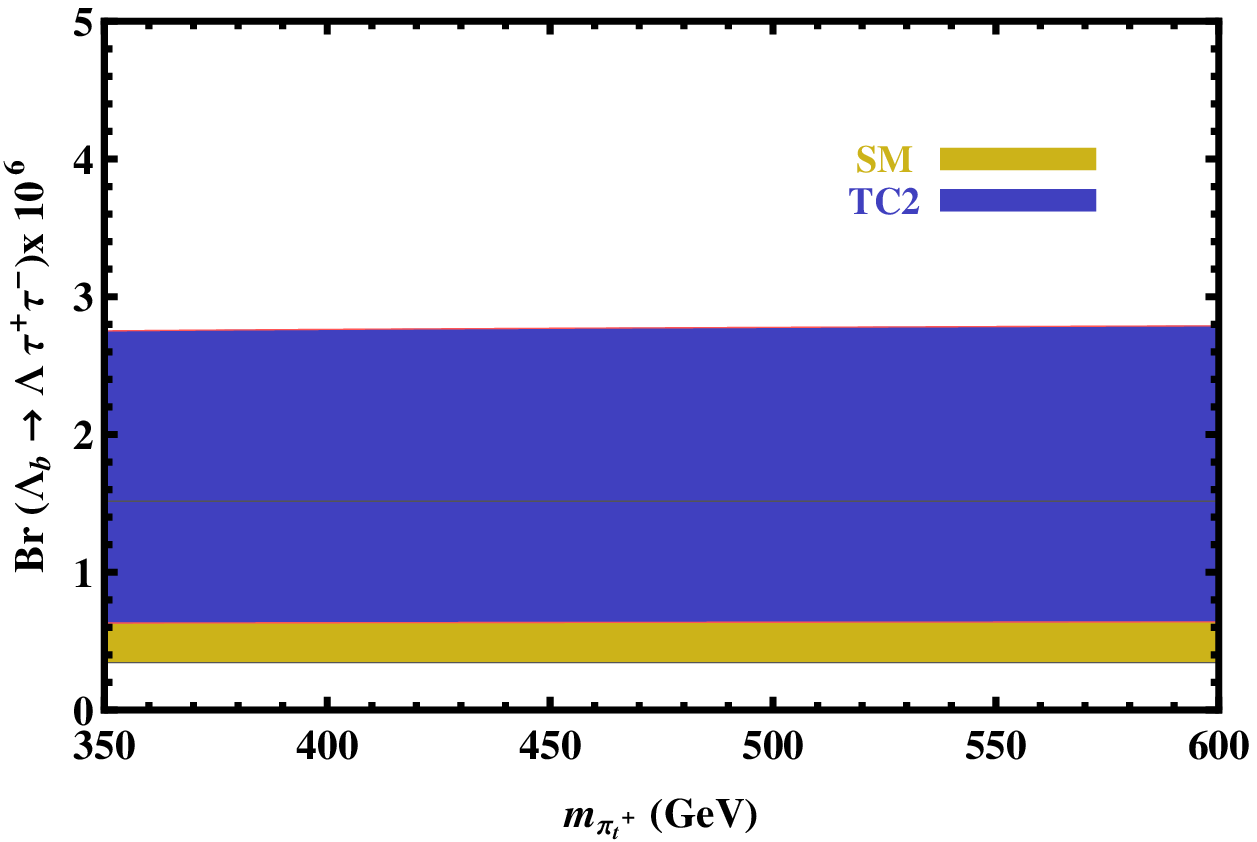,width=0.45\linewidth,clip=}
\end{tabular}
\caption{The same as figure 7 but for $\Lambda_b
\rightarrow \Lambda \tau^{+} \tau^{-}$ transition.}
\end{figure}
\begin{figure}[ht!]
\centering
\begin{tabular}{cc}
\epsfig{file=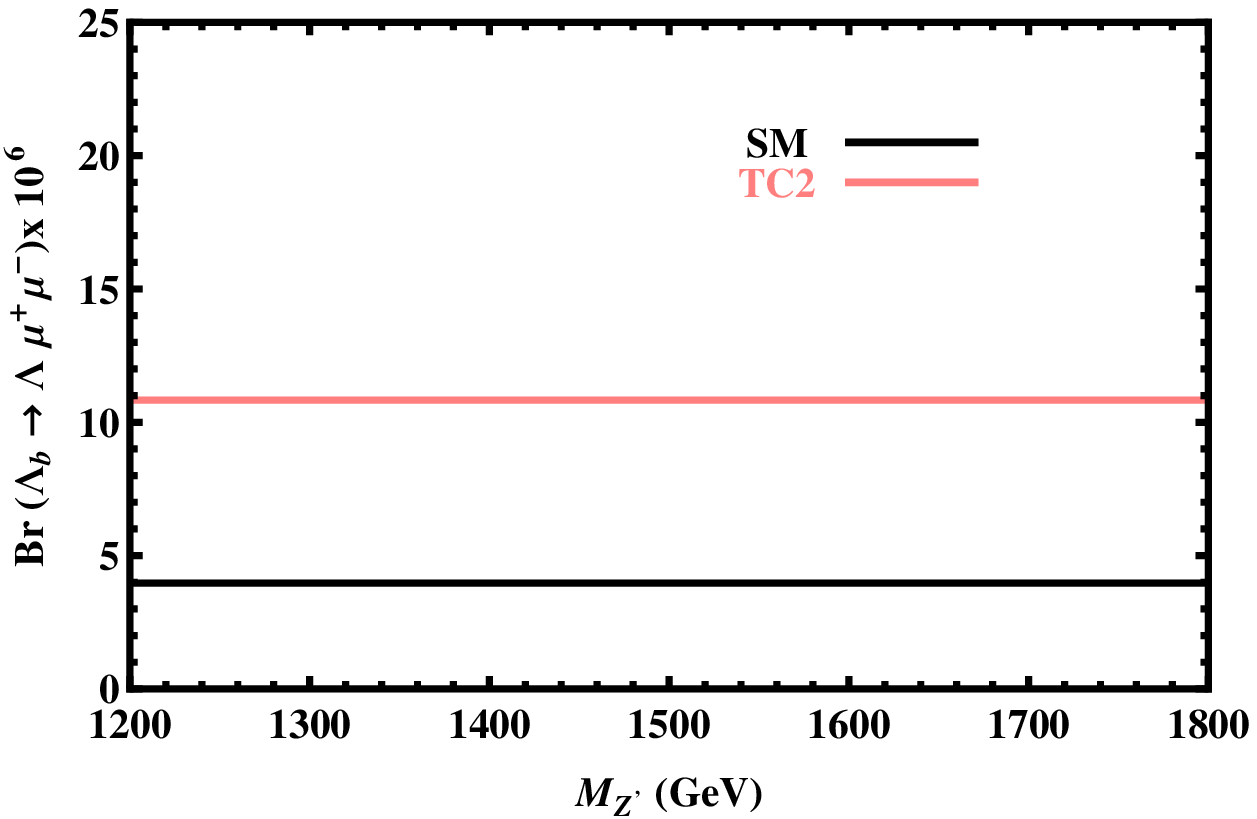,width=0.45\linewidth,clip=}&
\epsfig{file=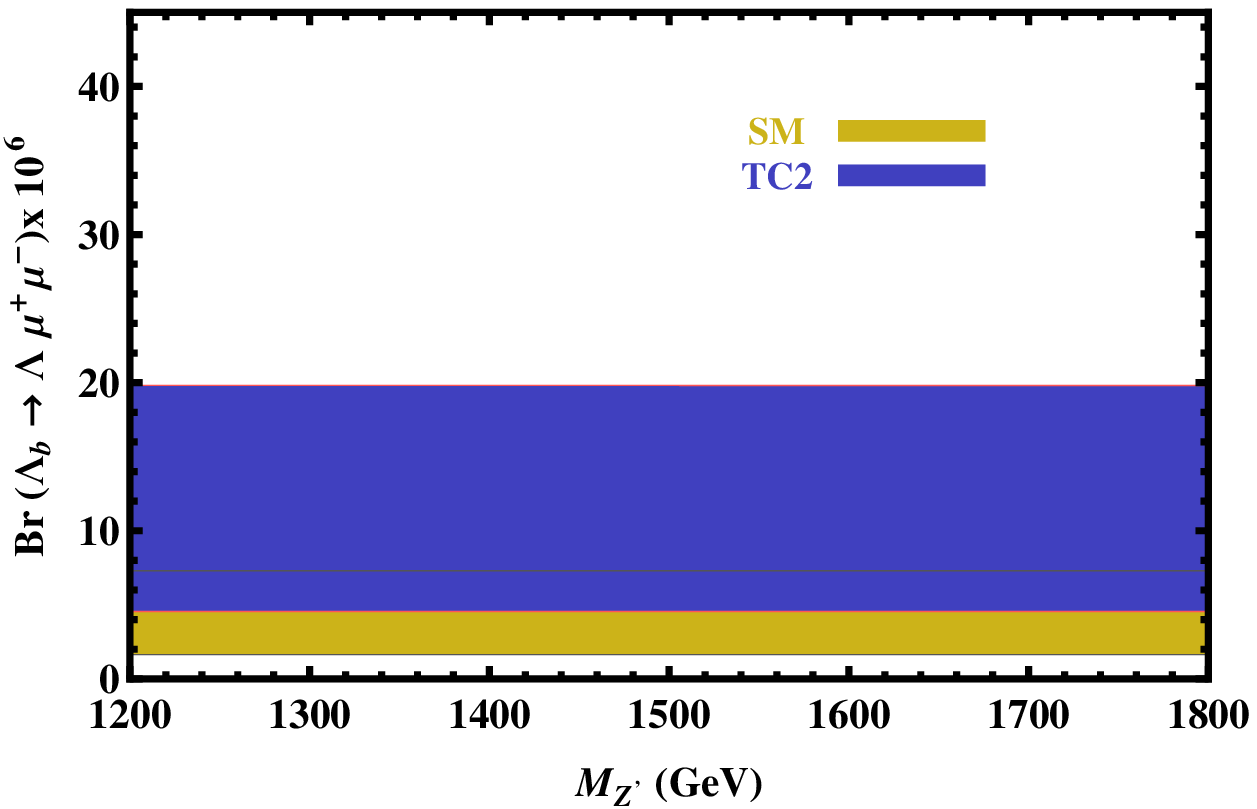,width=0.45\linewidth,clip=}
\end{tabular}
\caption{The same as figure 7 but with respect to $M_{Z^{\prime}}$.}
\end{figure}
\begin{figure}[ht!]
\centering
\begin{tabular}{cc}
\epsfig{file=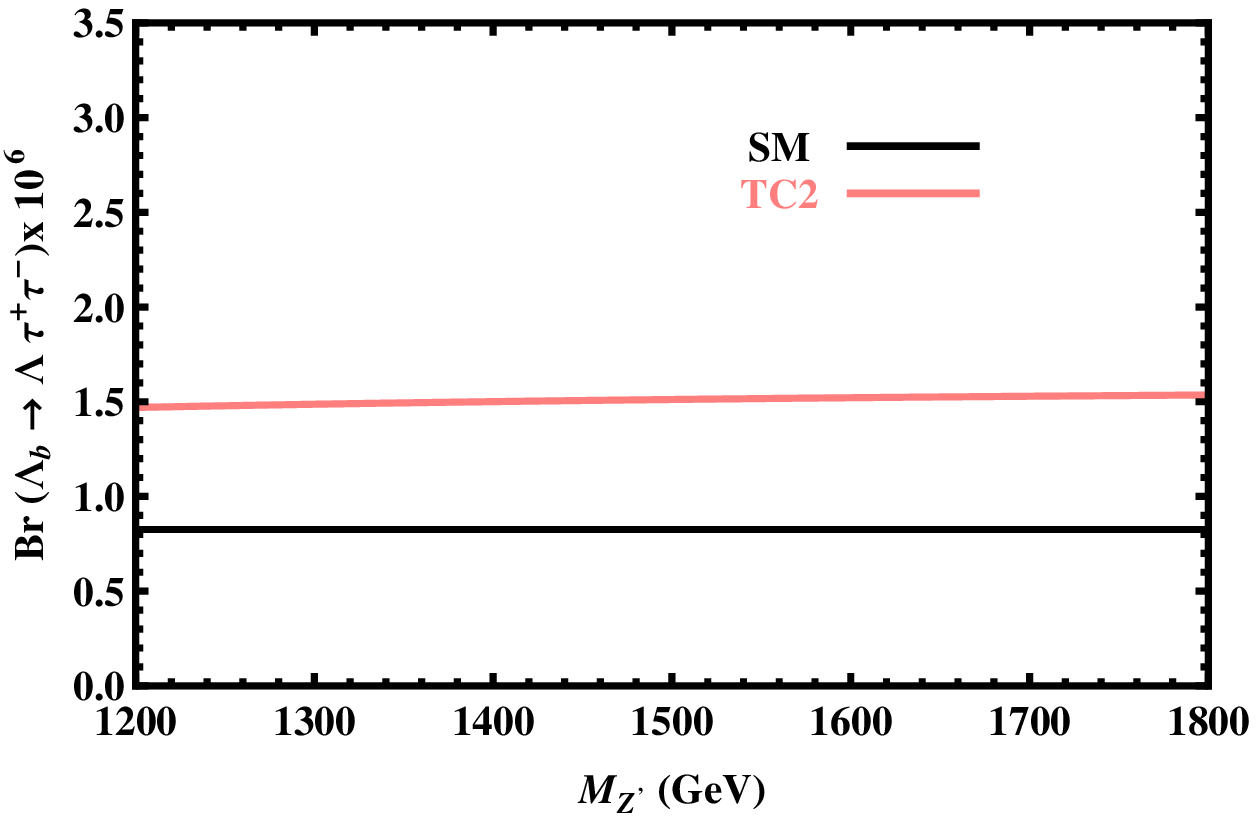,width=0.45\linewidth,clip=}&
\epsfig{file=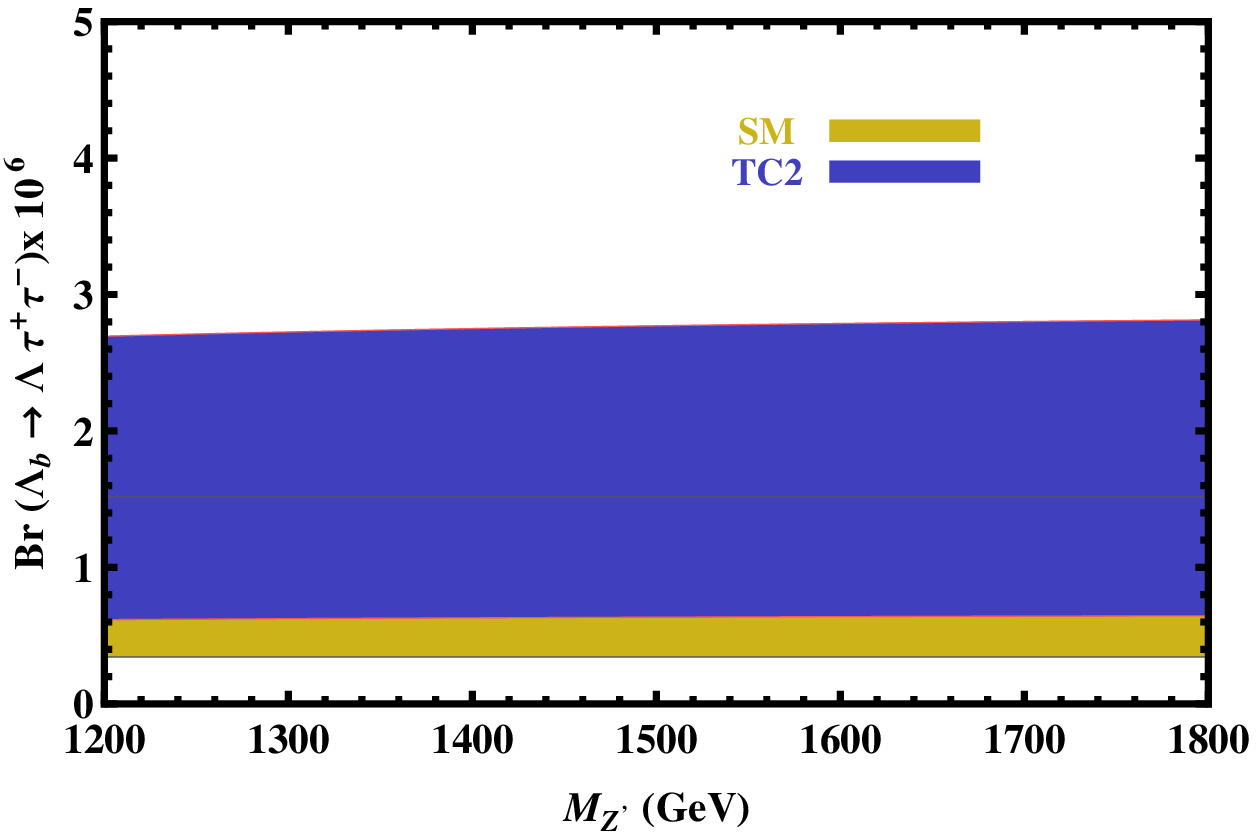,width=0.45\linewidth,clip=}
\end{tabular}
\caption{The same as figure 9 but for $\Lambda_b
\rightarrow \Lambda \tau^{+} \tau^{-}$ transition.}
\end{figure}
\subsection{The FBA}
The present subsection embraces our analysis on the lepton forward-backward asymmetry (${\cal A}_{FB}$) in both
the SM and TC2 models. The FBA
is one of the most important tools to investigate the NP beyond the SM and it is defined as 
\begin{eqnarray} {\cal A}_{FB} (\hat s)=
\frac{\ds{\int_0^1\frac{d\Gamma}{d\hat{s}dz}}(z,\hat s)\,dz -
\ds{\int_{-1}^0\frac{d\Gamma}{d\hat{s}dz}}(z,\hat s)\,dz}
{\ds{\int_0^1\frac{d\Gamma}{d\hat{s}dz}}(z,\hat s)\,dz +
\ds{\int_{-1}^0\frac{d\Gamma}{d\hat{s}dz}}(z,\hat s)\,dz}~. 
\end{eqnarray}
The dependences of the FBA on $q^{2}$, $m_{\pi_{t}^{+}}$ and $M_{Z^{\prime}}$ for the
 decay under consideration in both $\mu$ and $\tau$ channels are depicted in Figures 11-16. A quick glance
 at these figures leads to the following results
\begin{itemize}
\item the effects of the uncertainties of form factors on ${\cal A}_{FB}$ are smaller compared to the differential
branching ratio and branching ratio discussed in the previous figures.
\item In Figure 11 where the dependence of ${\cal A}_{FB}$ on $q^2$ at $\mu$ channel
is discussed, we see considerable differences between two models predictions at lower values of  $q^2$ for both
figures at left and right panels. At higher values of  $q^2$, two models have approximately the same predictions.
In $\tau$ case (Figure 12), two models have roughly the same results.
\item In the case of ${\cal A}_{FB}$ in terms of $m_{\pi_{t}^{+}}$ and $\mu$ channel, the uncertainties of
form factors end up in some common regions between two models predictions. In the case of $\tau$ channel,
the difference between two models results exists even considering the uncertainties of form factors.
\item In the case of ${\cal A}_{FB}$ on $M_{Z^{\prime}}$ and $\mu$ channel, we see a small difference between the SM 
and TC2 models predictions when the central values of form factors are considered. Taking into account uncertainties
of form factors causes some intersections between two models predictions. In the case of $\tau$ (Figure 16), we see
considerable discrepancies between two models predictions which can not be killed by the uncertainties of
form factors.
\item The ${\cal A}_{FB}$ is sensitive to $q^2$ for both leptons. The ${\cal A}_{FB}$ is also sensitive to 
$M_{Z^{\prime}}$ only for the case of $\tau$. However, this quantity remains roughly unchanged with respect to
changes in $m_{\pi_{t}^{+}}$ for both lepton channels as well as with respect to $M_{Z^{\prime}}$ only 
for $\mu$ channel.
\end{itemize}
\begin{figure}[ht!]
\centering
\begin{tabular}{ccc}
\epsfig{file=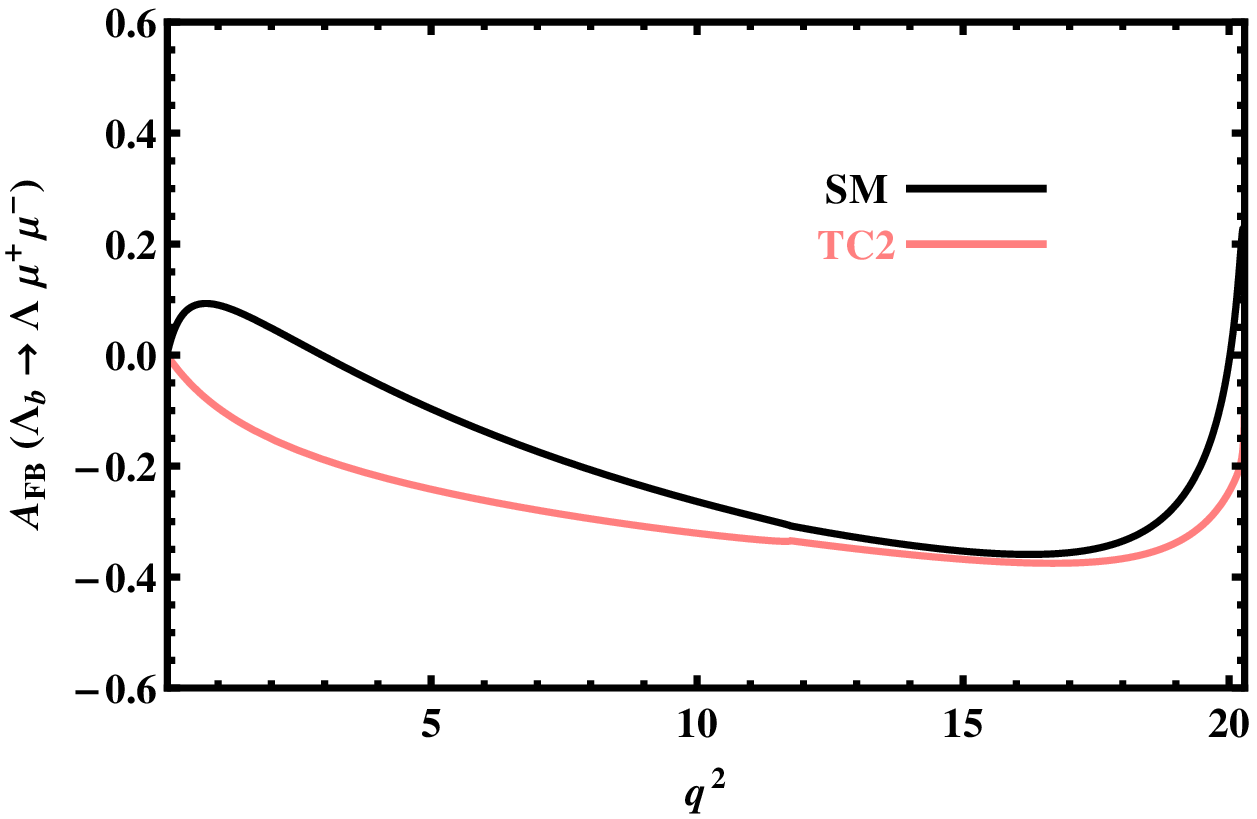,width=0.45\linewidth,clip=}&
\epsfig{file=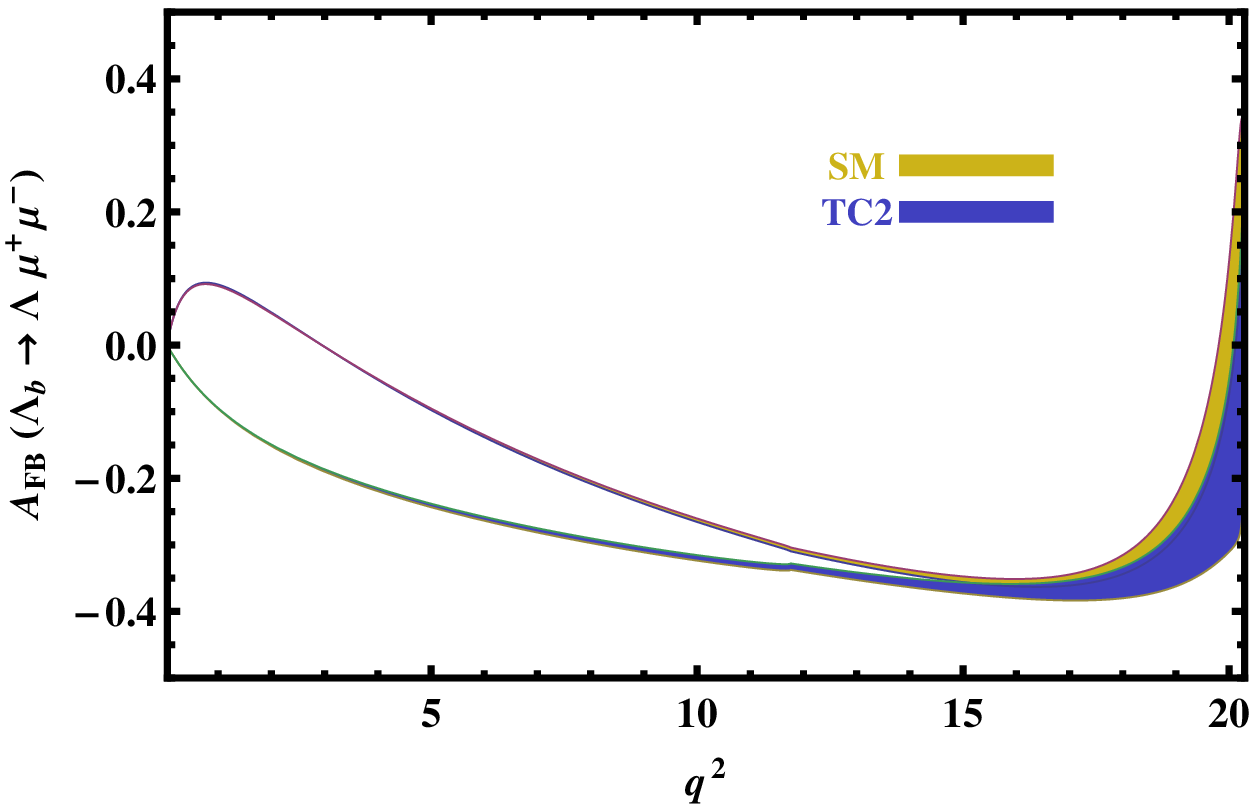,width=0.45\linewidth,clip=}
\end{tabular}
\caption{The dependence of the FBA on $q^{2}$ for
 $\Lambda_b\rightarrow \Lambda \mu^{+} \mu^{-}$ decay channel in the SM and TC2 models using the central 
values of the form factors (left panel) and form factors with their uncertainties (right panel).}
\end{figure}
\begin{figure}[ht!]
\centering
\begin{tabular}{ccc}
\epsfig{file=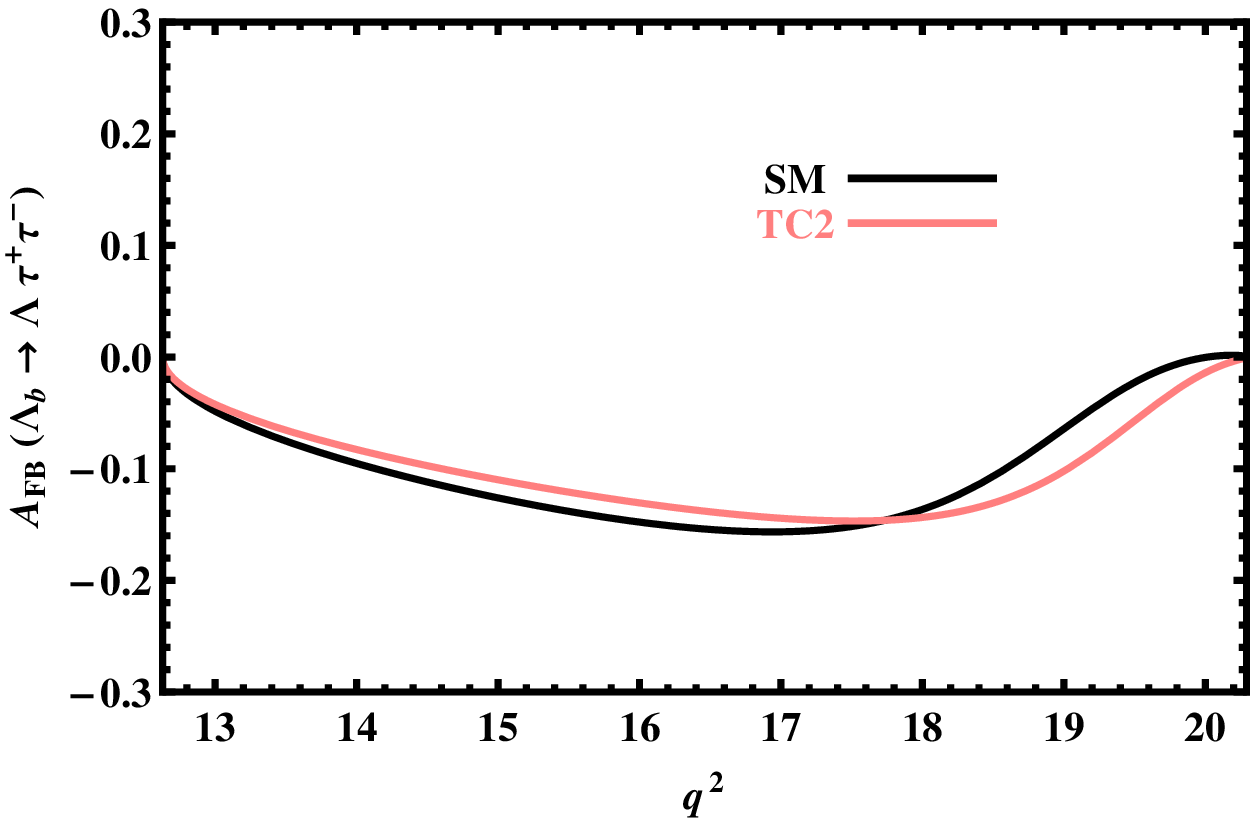,width=0.45\linewidth,clip=}&
\epsfig{file=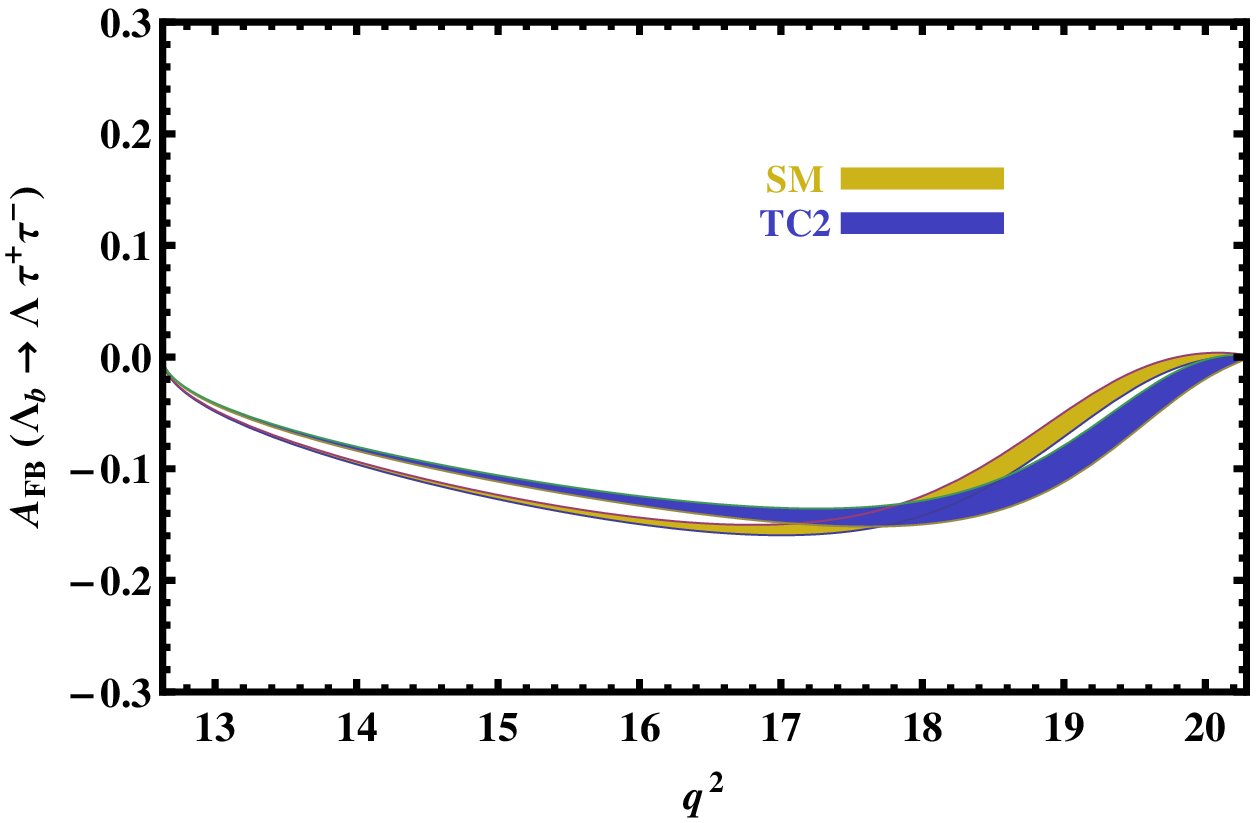,width=0.45\linewidth,clip=}
\end{tabular}
\caption{The same as figure 11 but for $\Lambda_b
\rightarrow \Lambda \tau^{+} \tau^{-}$ decay channel.}
\end{figure}
\begin{figure}[ht!]
\centering
\begin{tabular}{ccc}
\epsfig{file=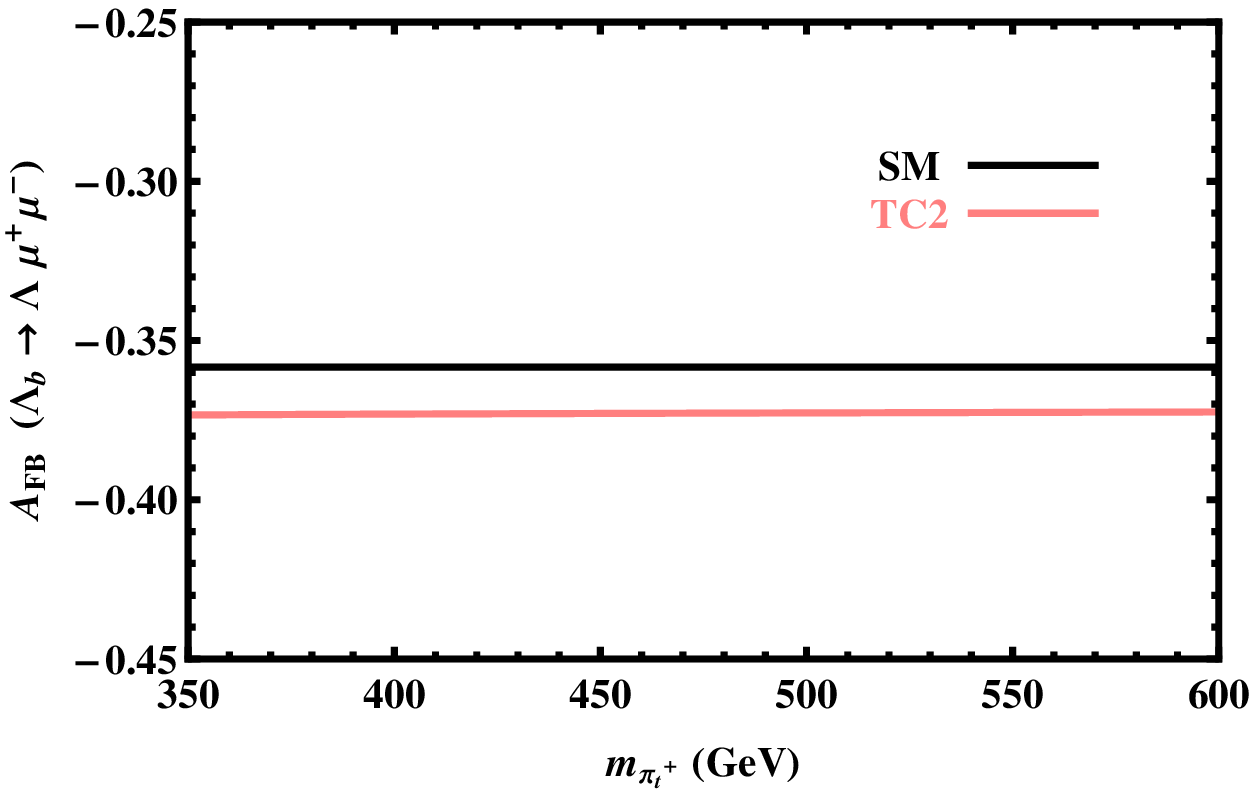,width=0.45\linewidth,clip=}&
\epsfig{file=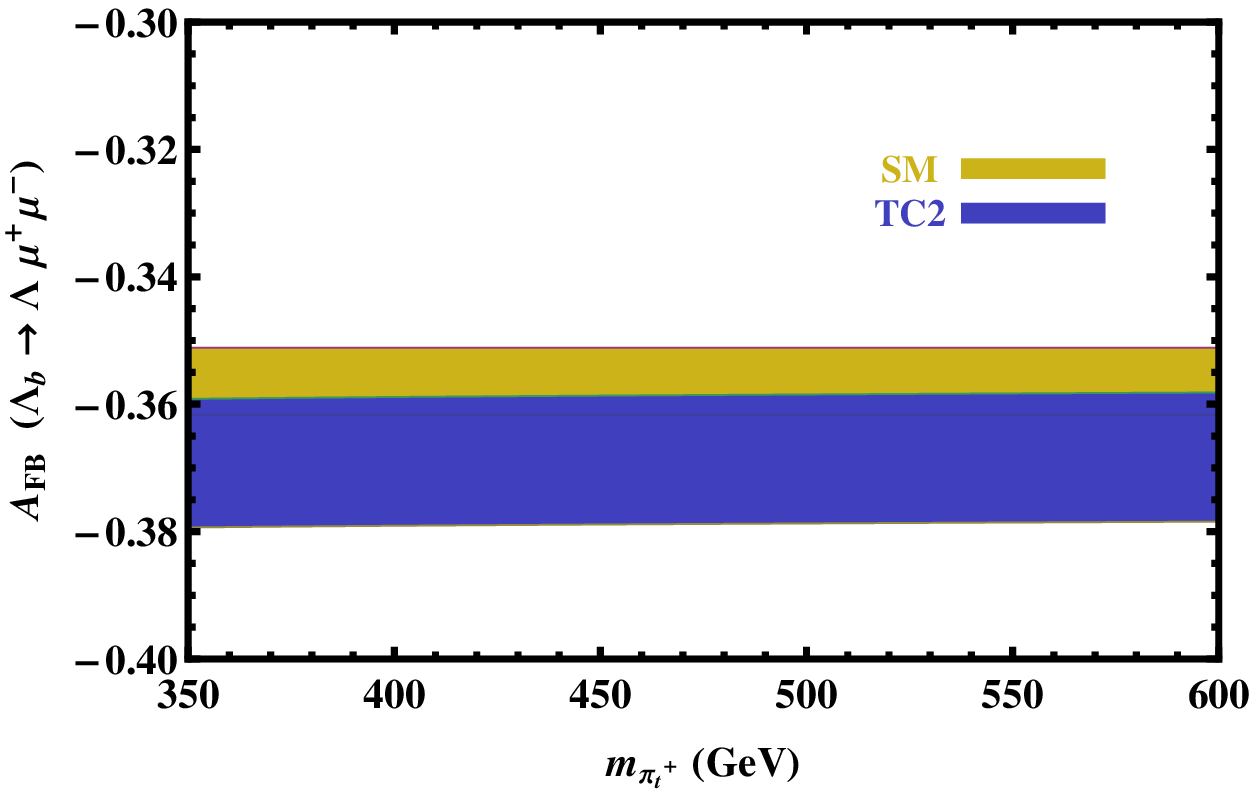,width=0.45\linewidth,clip=}
\end{tabular}
\caption{The same as figure 11 but with respect to $m_{\pi_{t}^{+}}$}
\end{figure}
\begin{figure}[ht!]
\centering
\begin{tabular}{ccc}
\epsfig{file=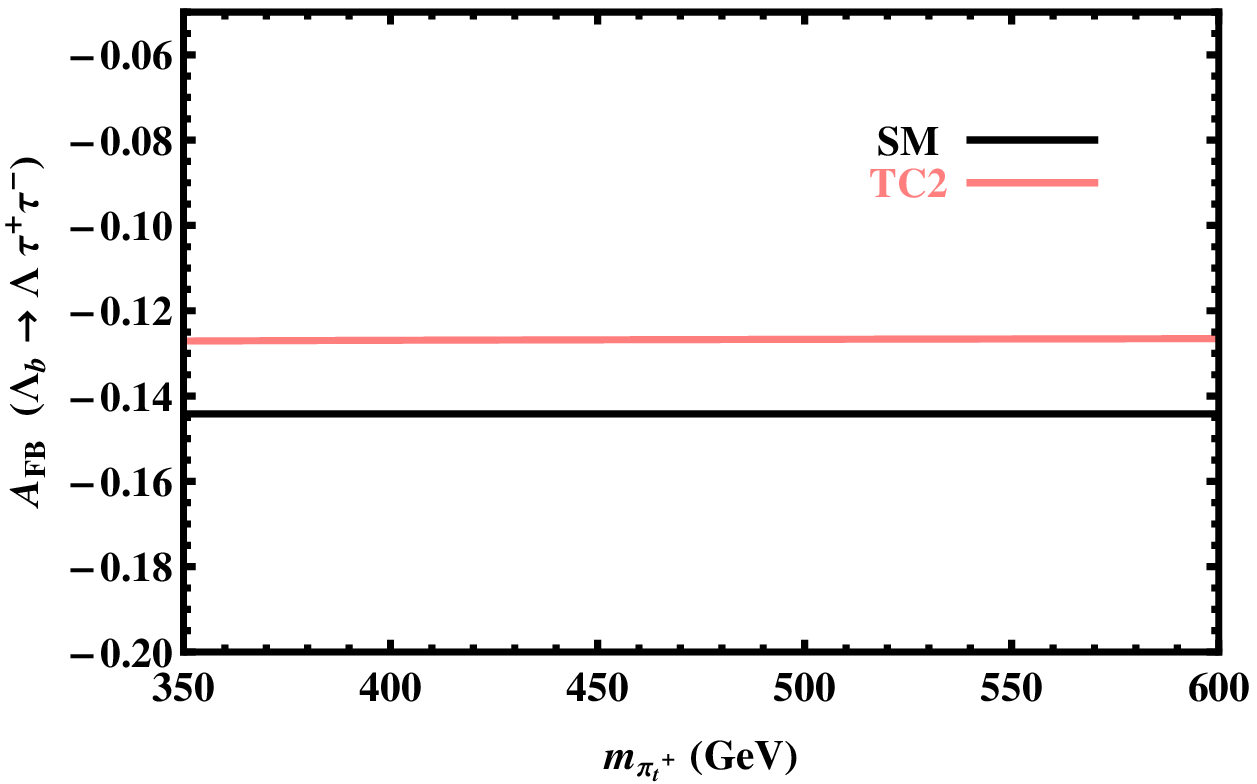,width=0.45\linewidth,clip=}&
\epsfig{file=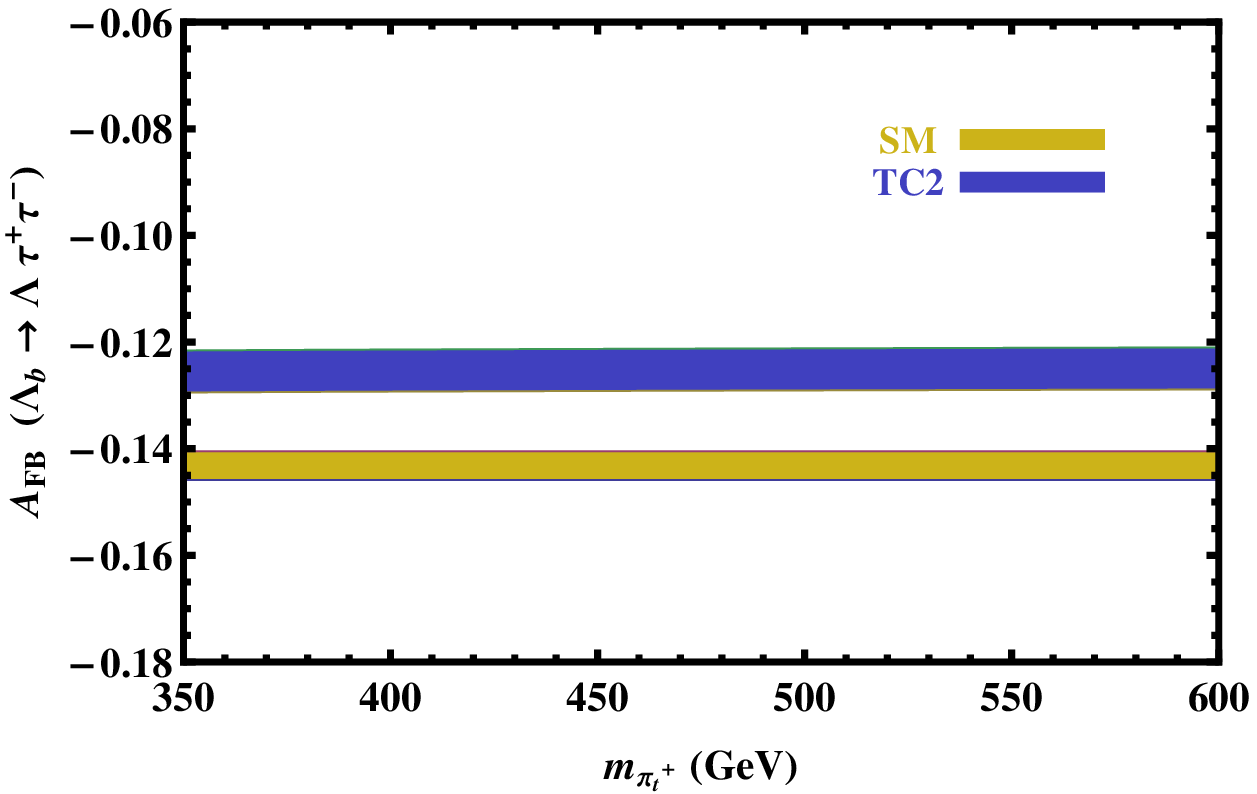,width=0.45\linewidth,clip=}
\end{tabular}
\caption{The same as figure 13 but for $\Lambda_b
\rightarrow \Lambda \tau^{+} \tau^{-}$ decay channel.}
\end{figure}
\begin{figure}[ht!]
\centering
\begin{tabular}{cc}
\epsfig{file=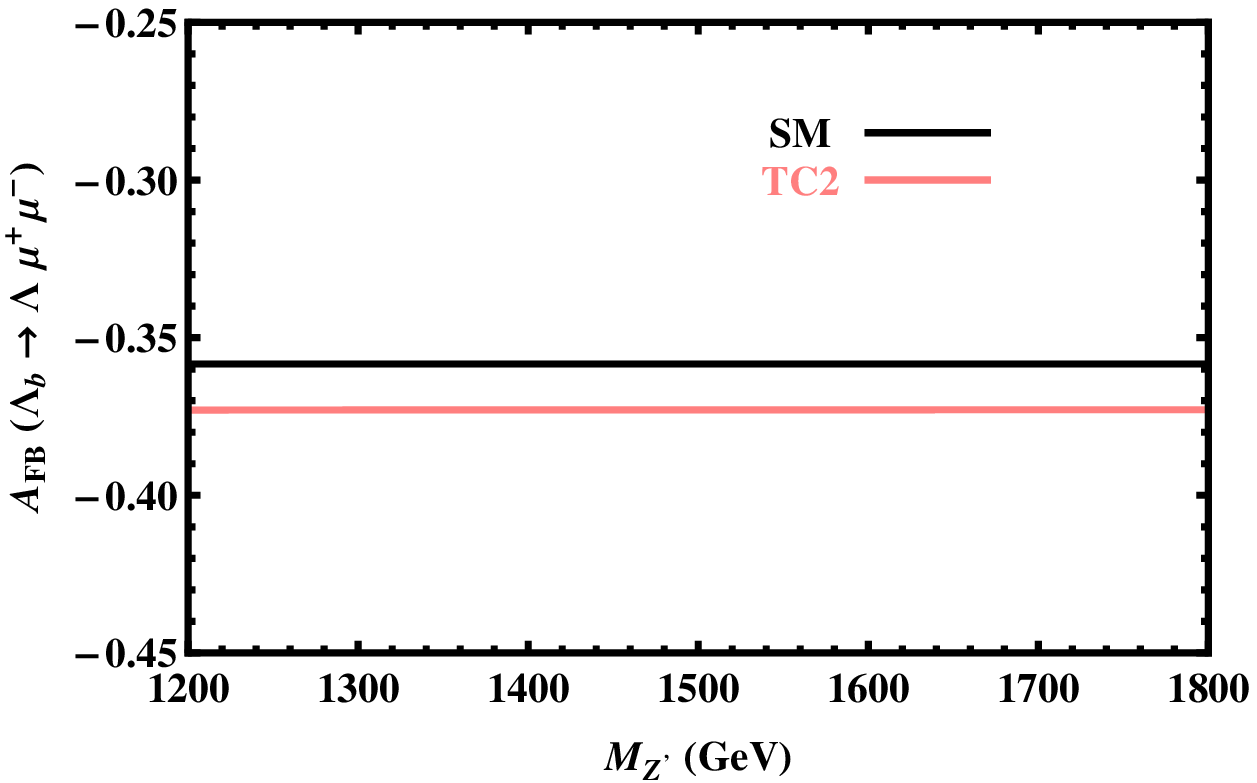,width=0.45\linewidth,clip=}&
\epsfig{file=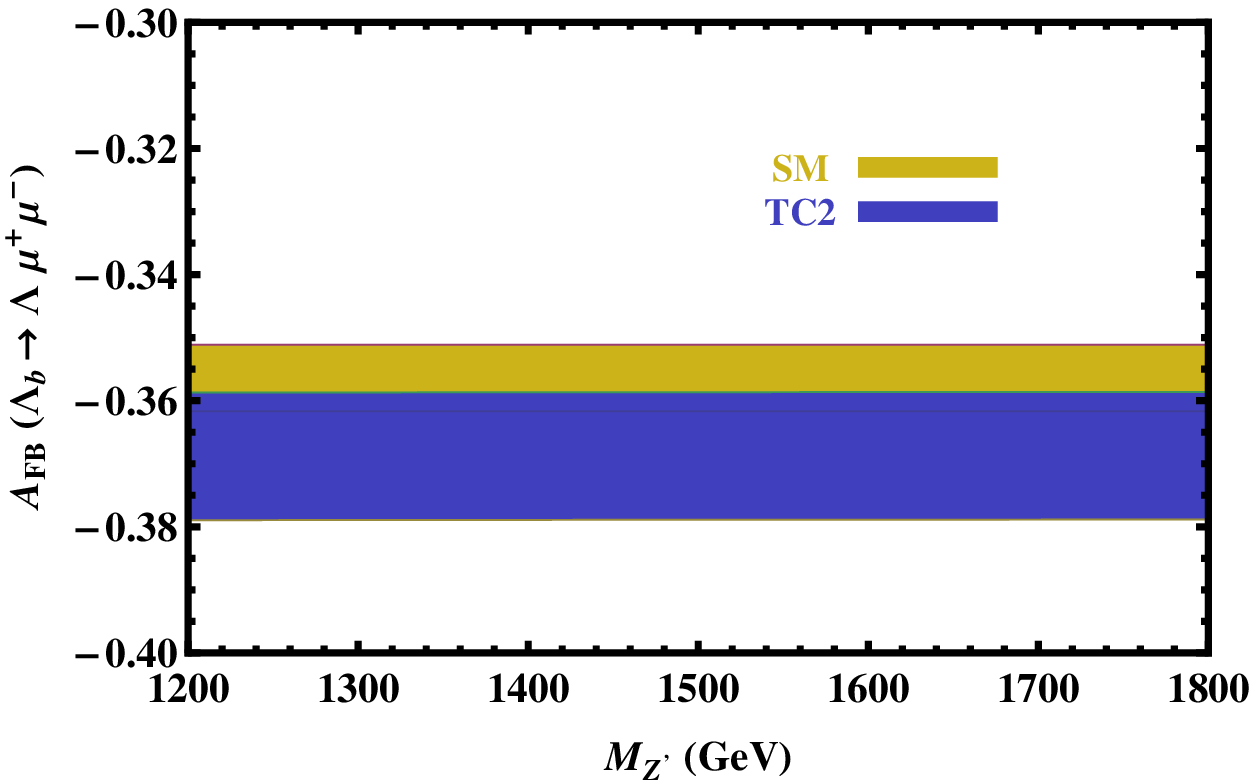,width=0.45\linewidth,clip=}
\end{tabular}
\caption{The same as figure 11 but with respect to $M_{Z^{\prime}}$.}
\end{figure}
\begin{figure}[ht!]
\centering
\begin{tabular}{cc}
\epsfig{file=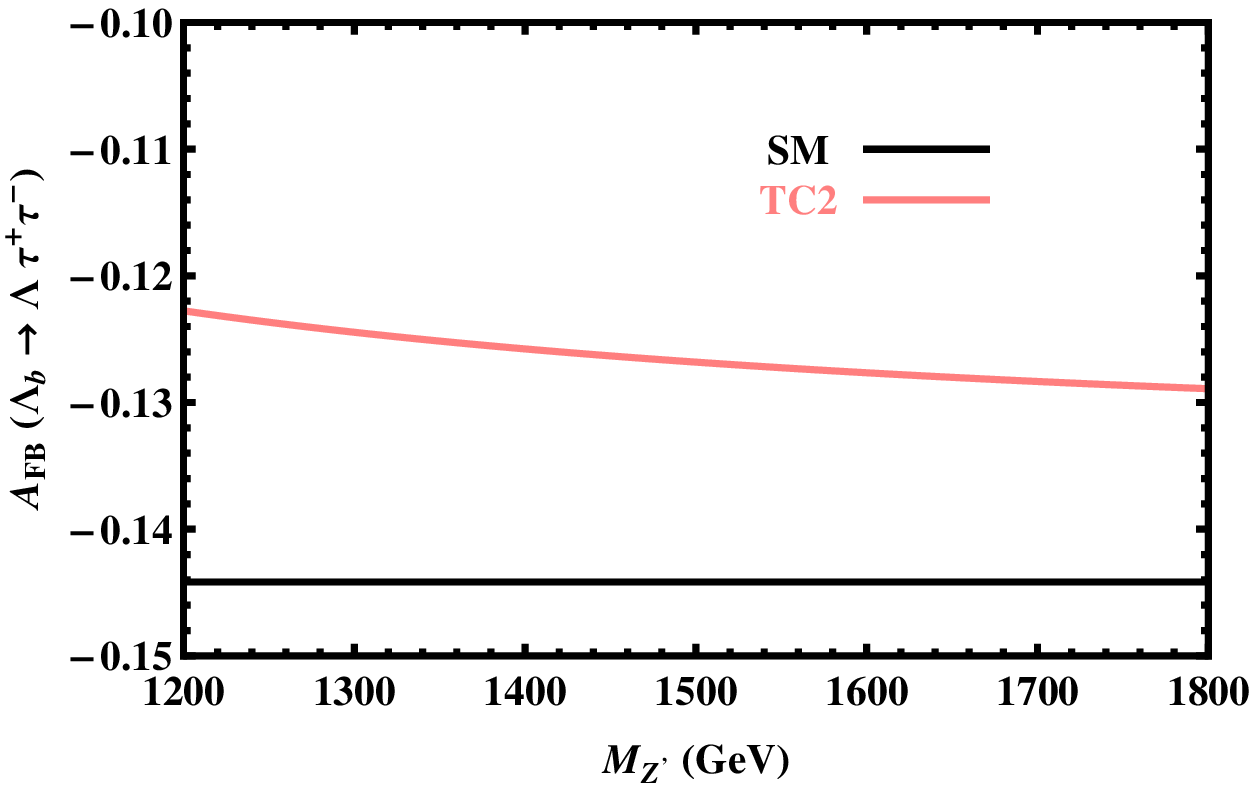,width=0.45\linewidth,clip=}&
\epsfig{file=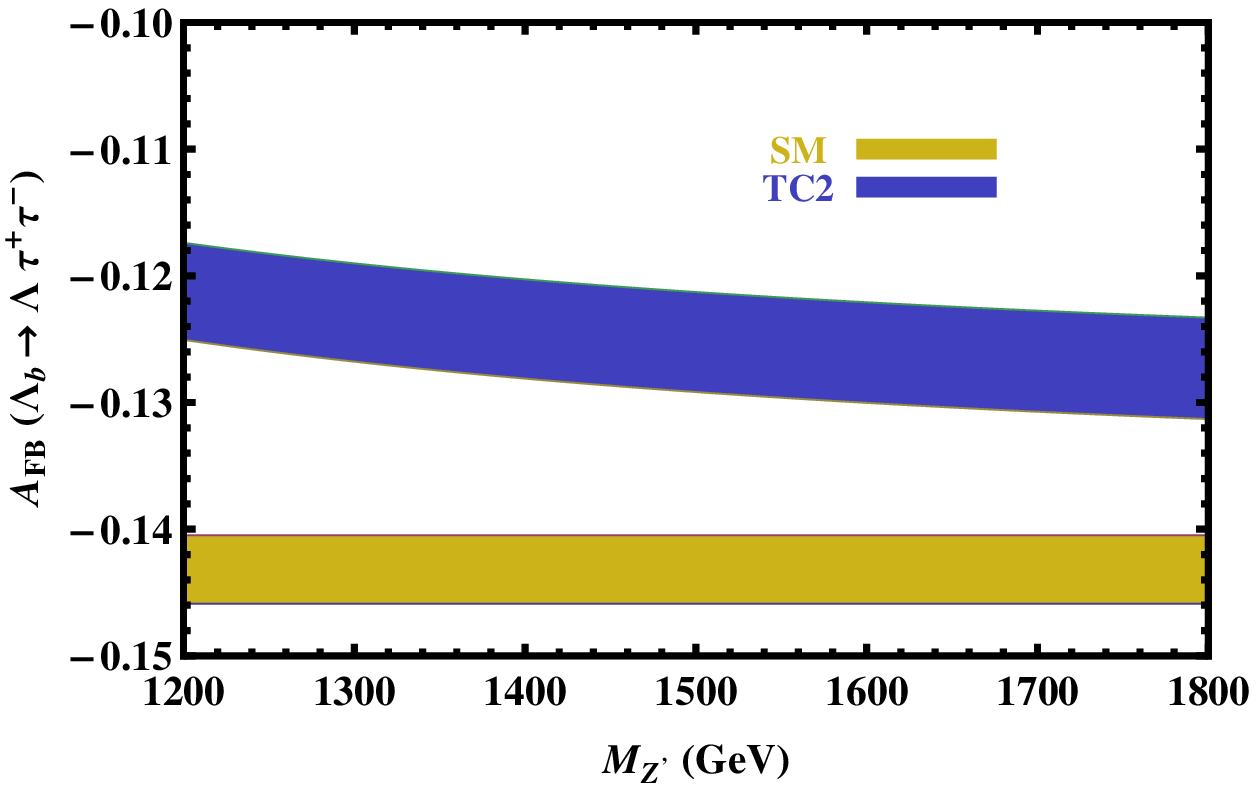,width=0.45\linewidth,clip=}
\end{tabular}
\caption{The same as figure 15 but for $\Lambda_b
\rightarrow \Lambda \tau^{+} \tau^{-}$ decay channel.}
\end{figure}
\section{Conclusion}
In the present work, we have performed a comprehensive analysis of the baryonic FCNC $\Lambda_b
\rightarrow \Lambda \ell^{+} \ell^{-}$ channel both in the SM and TC2 scenarios. In particular, 
we discussed the sensitivity of the differential branching ratio, branching ratio and lepton FBA on $q^2$ and 
the model parameters $m_{\pi_{t}^{+}}$ and $M_{Z^{\prime}}$ using the form factors calculated via light cone
QCD sum rules as the main inputs.  We saw overall considerable differences between two models' predictions,
which can not be totally  killed by the uncertainties of form factors as the main sources of errors. 
However, the existing experimental data provided by the CDF and LHCb Collaborations in the case of differential
branching ratio with respect to $q^2$ are very close to the SM results approximately in all ranges of $q^2$ when the errors of the form factors are considered. 
Only in some intervals of $q^2$,  the experimental data on the differential branching ratio  lie in the intervals predicted by the TC2 model within the errors. 
 From the experimental side, we think we should have more data on 
different physical quantities defining the decay under consideration at different lepton channels
as well as different baryonic and mesonic processes. This will help us in searching for NP effects 
especially those in the TC2 model as an alternative EWSB scenario to the Higgs mechanism.
\section{Acknowledgement}
 We would like to thank Y. Erkuzu for her contribution in early stages of the calculations. 

 
\end{document}